\documentclass{article}
\usepackage{naturetex}
\usepackage[dvipsnames]{xcolor}
\usepackage{siunitx}
\usepackage{soul}
\usepackage{cite}
\usepackage{ulem}
\usepackage{amsmath}
\usepackage{comment}

\usepackage[dvipsnames]{xcolor}

\usepackage[utf8]{inputenc}

\usepackage{multimedia}
\usepackage{tabularx}

\usepackage{xcolor}

\usepackage{afterpage}
\usepackage{graphicx} %The mode "LaTeX => PDF" allows the following formats: .jpg  .png  .pdf  .mps
\graphicspath{{./PresentationPictures/}} %Where the figures folder is located
\usepackage{media9}

\begin{document}

\maketitle
{%\setstretch{1.0}
	% *** ABSTRACT ***
	\section*{Abstract}
    Active networks composed of biopolymers and motor proteins provide versatile biomimetic systems that have advanced active matter physics and deepened our understanding of cytoskeletal dynamics and self-organization under diverse stimuli. In these systems, 
    activity arises in aqueous solutions where motor proteins cross-link biopolymers and generate active stress driving the emergent network behavior.
    Here, we establish the active network in the form of a sessile, multi-component droplet on a substrate and investigate how evaporation influences its dynamics. We focus on how mass loss and compositional changes in the droplet reshape the behavior of the active suspension.
    We show that capillary and Marangoni flows drive the self-organization of microtubules into a distinctive radial arrangement within the droplet. The cross-linking ability of motor proteins gives rise to a striking non-monotonic wetting behavior, where the extensile stresses generated by the motor proteins strongly affect the characteristic timescale of the contact-line retracting and subsequent expansion. 
Using a combined experimental and theoretical approach, we demonstrate the crucial role of crosslinking in evaporating microtubule networks, and explain how active stresses together with evaporation-induced flows govern the dynamics of reconstituted microtubule systems and their wetting behavior.\\
Evaporating droplets have recently attracted significant attention in the scientific community, and the findings of the setup presented in this study 
can have broad implications, ranging from self-organization and mechanical pattern formation in biological systems to questions about the origin of life.
}% setstretch

% *** INTRO ***
\newpage

\section*{Introduction}
   Cytoskeletal assemblies, in the form of microtubule networks, drive vital cellular processes such as intracellular cargo transport \cite{franker2013microtubule}, cell guidance during migration \cite{small2002microtubules, watanabe2005regulation}, and play a crucial role in mechanical stability \cite{brouhard2018microtubule}, cell morphology \cite{kelliher2019microtubule}, and cell division \cite{scholey2003cell}.
These functions result from the self-organization of microtubules and motor proteins which, through interactions at the molecular scale, give rise to the system’s large-scale emergent behavior.
Examples of such self-organization, both in-vivo and in-vitro, have been demonstrated in numerous studies over the past decades, using both experimental and theoretical approaches \cite{Alper, nedelec1997self, surrey2001physical, sanchez2011cilia, sanchez2012spontaneous, Goldstein, MONTEITH2016}.

A pivotal experimental system to investigate the self-organization of cytoskeleton components consists of a reconstituted in-vitro assay made of suspensions of microtubules and multi-headed kinesin motors \cite{sanchez2012spontaneous}. These suspensions are driven out of equilibrium by kinesin motors, which bind to the microtubules and move along them, exerting active stresses during their stepping motion, powered by a continuous supply of adenosine triphosphate (ATP) in solution. In such in-vitro assays, the activity of the motors is enhanced by organizing the microtubules into bundles through the addition of depleting agents such as polyethylene glycol (PEG), which induces effective attractive interactions between randomly oriented microtubules via entropic forces~\cite{asakura1958interaction,needleman2004synchrotron,sanchez2012spontaneous}. Specifically, PEG acts as a depletion agent, bundling the microtubules and promoting kinesin motor binding, thereby enhancing force generation through their active motion along the filaments. Under this arrangement, the motors generate active forces that lead to spatial displacements of the microtubules and bending of the bundles at larger length scales, as well as to emergent behaviors such as the formation of microtubule asters, vortices, or nematic structures in 2D and 3D~\cite{surrey2001physical,nedelec1997self,sanchez2012spontaneous,strübing2020wrinkling}.

Suspensions of microtubules and kinesin motors have been studied in various contexts, most notably when confined at the interface between two fluids~\cite{sanchez2012spontaneous}, or within channels with rigid boundaries~\cite{opathalage2019self,strübing2020wrinkling,hardouin2019reconfigurable}. Of particular interest is the influence of the external environment on such suspensions, especially how they respond to mechanical cues. Indeed, gliding assay experiments have shown that applying external stress can influence motor activity and tune the spatio-temporal distribution of the internal driving forces leading to the emergent behaviour of the system \cite{Akira2019}. Several in-vitro experimental setups have been developed to investigate the response of reconstituted active cytoskeletal networks to externally applied stimuli. These include patterned surfaces, varying confinement geometries, and interactions at water/oil interfaces~\cite{ross2019controlling,opathalage2019self,Guillamat2017}.
At fluid interfaces, microtubule–kinesin motor suspensions exhibit fascinating behaviors, such as active turbulence, characterised by the continuous creation and annihilation of defects in the microtubule orientation field~\cite{tan2019topological}, and defect ordering driven by changes in friction with the surrounding fluid~\cite{decamp2015orientational}.
Theoretical attempts to understand these phenomena include analyses of the dynamics of a nematic order parameter, representing the orientation of force-generating microtubules, coupled to a velocity field and the resulting instabilities that lead to the formation of flow patterns and defects~\cite{giomi2013defect,mechbio,mmykf9x3,C6SM00812G}; the ordering of defects in the presence of friction with the external environment~\cite{srivastava2016negative,sultan2022quadrupolar,putzig2016instabilities,PhysRevLett.125.218004}; and the transition to flow in confined geometries~\cite{strübing2020wrinkling,Varghese,PRXLife.1.023008}.

Here we study the behaviour of these active suspensions of microtubules and kinesin motors when they are in a liquid droplet deposited on a substrate and evaporate in the environment. 
Previous studies reported the dynamics of active microtubule-kinesin network inside droplet-like compartments, namely water-in-oil droplets \cite{Ahmad2021}, lipid vesicles \cite{keber2014topology} and water/water phase separation (w/wPS) droplet composed of a Dextran-PEG mixture \cite{Sakuta}. 
On the other hand, the droplet-like confinements mentioned above are immersed in a bulk aqueous solution  and therefore present a liquid–liquid interface.
The intriguing aspect of the active droplets we investigate is that they are not confined in a close experimental chambers with aqueous solution. They rather exhibit a dynamic liquid–vapor interface, across which the mass of the suspension changes over time due to evaporation. Droplet evaporation induces dynamic changes in the geometry of confinement, leading to outward capillary flows \cite{deegan1997}. However, preferential evaporation can generate compositional gradients that—even in passive systems—drive inward Marangoni flows and exert mechanical stresses on the suspended constituents. The aggregation of colloids into ordered or disordered structures under such flows has been the subject of extensive studies~\cite{deegan1997,pearson_1958,berg_boudart_acrivos_1966,Cloot1990,marin2011order,marin2012building,marin2019solutal}. 
In contrast, to the best of our knowledge, the behavior of an active evaporating droplet deposited on a substrate and composed of microtubules and kinesin motors has not been reported so far. We first demonstrate that evaporation-induced flows, associated with the Marangoni effect, drive the self-organization of microtubule filaments into a regular radial structure inside the droplet. Then, we study the long-term wetting behavior of the microtubule–kinesin droplet under evaporation and the effects of different constituents such as PEG and ATP on its dynamics. 
We perform experiments and use lubrication theory to understand these observations. Using a minimal lubrication model, we show that the combination of the Marangoni effect and the extensile activity of microtubule–kinesin networks generates a fluid flow that drives the initial retraction and later expansion of the droplet and that activity is able to accelerate the dynamics by an order of magnitude.

Evaporating microdroplets have recently attracted the attention of the scientific community as the unique properties of the air–water interface can drive chemistry that is unlikely to occur in bulk. The characterisation of such set up can have major implications across a wide range of fields including self-organization, mechanical pattern formation, and the origin of life \cite{Life_deal2021,Life_Dobson2000,LifeTuck2002}. In this study we analyse the response of a biological network inside a microdroplet to the evaporation kinetics of its compartment and how out-of-equilibrium ATP-driven activity influences its dynamics.

\section*{Results and discussion}
\begin{figure}[ht!]
	\begin{center}
		\includegraphics[width=\textwidth]{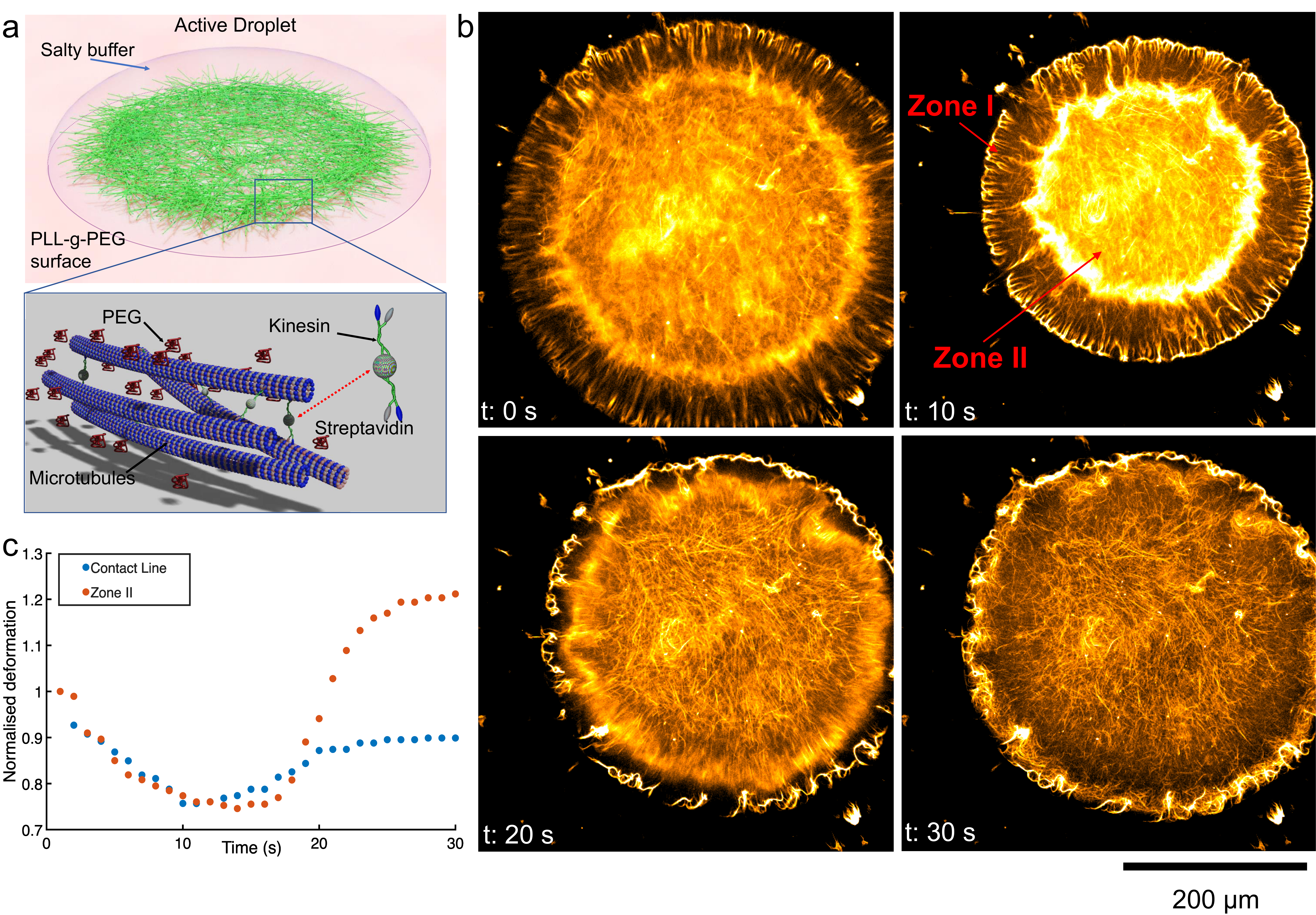}
	\end{center}
	\caption{Dynamics of active microtubule network inside an evaporating droplet. a) Schematic representation of the experimental setup. Microtubule active network embedded into a droplet of a mixture of salty buffer and PEG on a PLL-g-PEG functionalised substrate. The inset shows the biological building blocks constituting the network, namely microtubules arranged into bundles due to PEG depletion force and kinesin-streptavidin motor clusters. b) Micrographs showing the emergent behaviour of the self-organizing active network during the droplet evaporation over a time of 30 s. The pattern can be described by two zones marked as zone I, and zone II, which we refer to as the \textit{radial corona} and the \textit{dense network}, respectively. c) Dynamics of the contact line of the droplet and isotropic network during evaporation.
	\label{fig:fig1}}
    \end{figure}

A droplet extracted from an active mixture composed of stabilized microtubules, kinesin-1 motor proteins organized in clusters, PEG, and ATP (prepared as described previously~\cite{strübing2020wrinkling}) was deposited onto a PLL-g-PEG–functionalised glass slide (see Materials and Methods for details). A schematic representation of the system is shown in Figure~\ref{fig:fig1}(a). 
The presence of PEG in the mixture promoted the formation of self-assembled microtubule bundles. The average length of individual microtubule filaments was approximately 15~$\mu$m, but the bundles they formed exceeded this length by an order of magnitude. The microtubules were fluorescent labeled to enable visualisation. The addition of kinesin-1 motor proteins facilitated cross-linking both within and between bundles (see inset in Figure~\ref{fig:fig1}(a)), organizing them into a network. These motor proteins move along the microtubules fueled by the energy of ATP hydrolysis, generating active (i.e., non-equilibrium) stresses within the network, as shown in a previous study \cite{nasirimarekani2022}.
The network was visualised by fluorescence microscopy, as shown in Figure~\ref{fig:fig1}(b) and its continuous evolution was recorded at several time points during the droplet evaporation into the ambient atmosphere, which began immediately after depositing the mixture onto the glass slide. In this configuration, the microtubules were subject not only to the forces exerted by the motor proteins but also to the evaporation-driven flows inside the droplet, namely capillary and Marangoni flows. This combination led to a unique pattern of network density and orientation of microtubule bundles as well as a distinctive combination of dynamic processes.

The visualisation started a few seconds after the deposition on the substrate, corresponding to $t = 0$~s in Figure~\ref{fig:fig1}(b).
 After depositing the droplet, we observed, near the contact line, an annular region in which radially oriented microtubule bundles were relatively uniformly distributed around the droplet. We refer to this structure as a \textit{radial corona}. Toward the center of the droplet, these gave way to isotropically and randomly oriented microtubule bundles that formed a \textit{dense network}.
Based on these observations, we divided the pattern formed by the microtubules into two distinct sectors, labeled zone I and zone II, according to the fluorescence signal (see Figure~\ref{fig:fig1}(b)) and corresponding to the annular region and the isotropically organized network, respectively, described above.
Qualitatively, the dynamics of the system is characterised by an initial contraction, followed by a spreading motion
of the droplet footprint. Movie S1 (Supplementary Information, SI) provides a comprehensive visualisation of this dynamics.
Specifically, after an initial period we observed the receding motion of the contact line (Figure~\ref{fig:fig1}(b) at $t = 10$~s). Notably,  we did not find any filament deposition at the contact line during this contraction. This is attributed to the 
PLL-g-PEG–functionalised glass substrate, which reduces nonspecific protein adsorption on the surface. We proved this by depositing a microtubule droplet on a non-functionalised substrate and we confirmed the accumulation of filaments at the contact line without functionalisation (See Figure S1 in SI). 

Interestingly, we did not observe the active network in the central region (zone II) collapsing into an aster structure, as has been reported in isotropic networks of microtubules and kinesin motors in closed chambers~\cite{nedelec1997self,Torisawa}.  Instead, the network continued to fill zone II and resisted the radial Marangoni flow likely due to its elastic cross-linked  feature~\cite{sanchez2012spontaneous,strübing2020wrinkling}. 
Afterwards, at $t\sim 20$~s the droplet spread to a larger
footprint diameter (Figure~\ref{fig:fig1}(b) at $t = 20-30 s$~s). 
 The 
 %annulus of radially 
 radial corona of oriented microtubule
bundles (zone I) gradually disappeared in this phase, the bundles from this region were advected into the contact line and the isotropic network (zone II) expanded through the entire droplet.

To understand the mechanism underlying the formation of the radial corona pattern as well as shrinking-spreading non-monotonic dynamics of the droplet contact line, we analysed the role of each constituent component by selectively removing them from the experimental setup.  
Although these events -- radial corona pattern formation and motion of contact line and enclosed network -- arise simultaneously in our experiments, we analyse and discuss them separately in the following sections for clarity.

\subsection*{Formation of the radial corona pattern }
\subsubsection*{Marangoni flow and development of the radial corona of microtubules}
\begin{figure}[h!]
	\begin{center}
		\includegraphics[width=\textwidth]{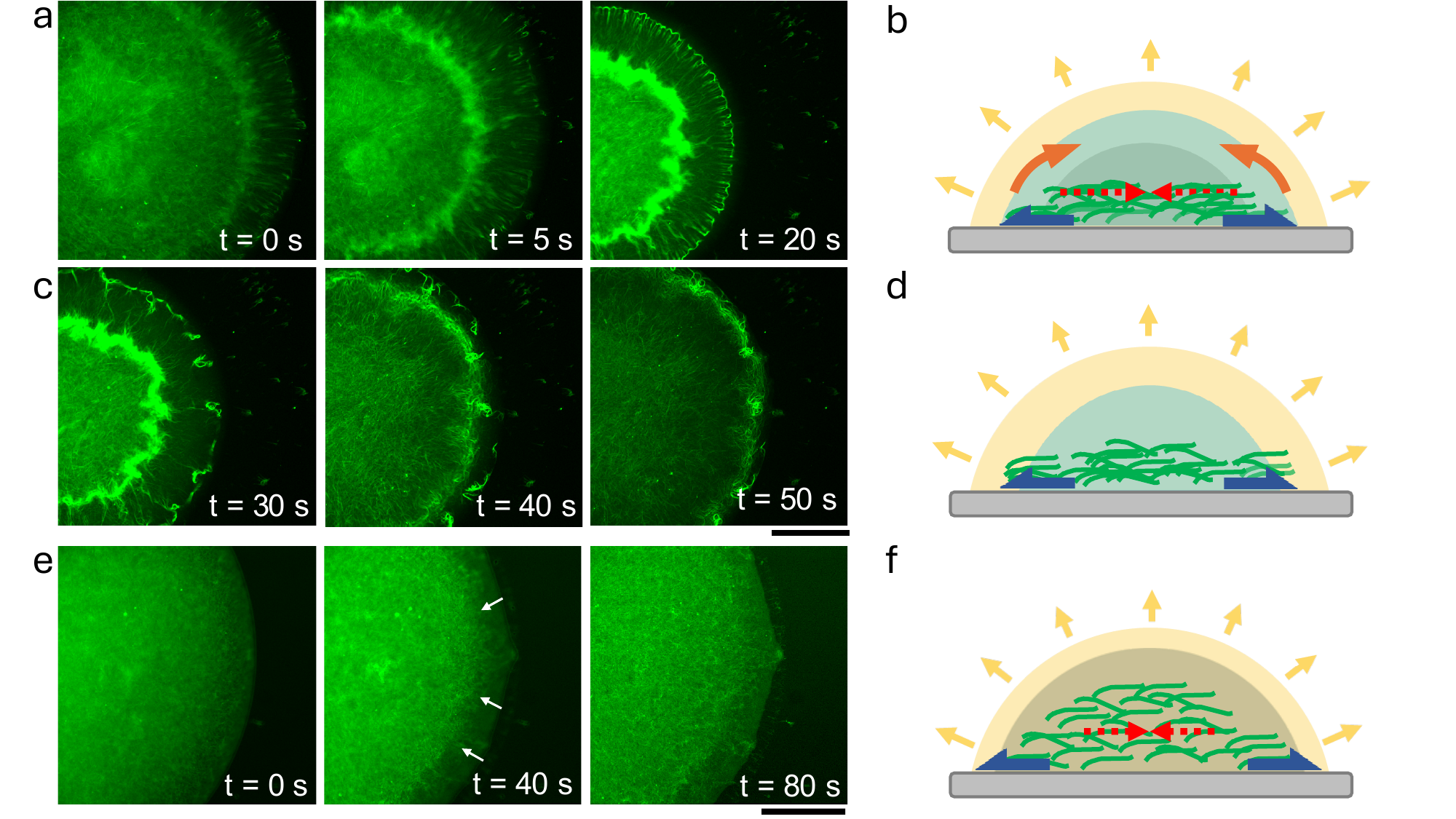}
	\end{center}
	\caption{Comparison of the dynamics of the active microtubule network in the presence (a–d) and absence (e–f) of PEG.
 (a)-(d): Micrographs and schematics of the droplet containing PEG for time interval of t = 0 -- 20 s, showing an inward contraction of the microtubule network and  the contact line at short time, and an outward extension at long times (t = 30 -- 50 s). In the schematics the yellow arrows represent the evaporation of the droplet, the orange arrows the Marangoni flow, the blue arrows the capillary flow and the red dashed arrows the network contraction force. (e) and (f): Micrographs and schematics of the active microtubule network in a droplet without PEG, showing a slight inward motion of the contact line and the contraction of the network (zone II, which is highlighted by white arrows). Scale bars: 200 $\mu$m. }
	\label{fig:fig2}
\end{figure}

\begin{figure}[ht!]
	\begin{center}
		\includegraphics[width=1\textwidth]{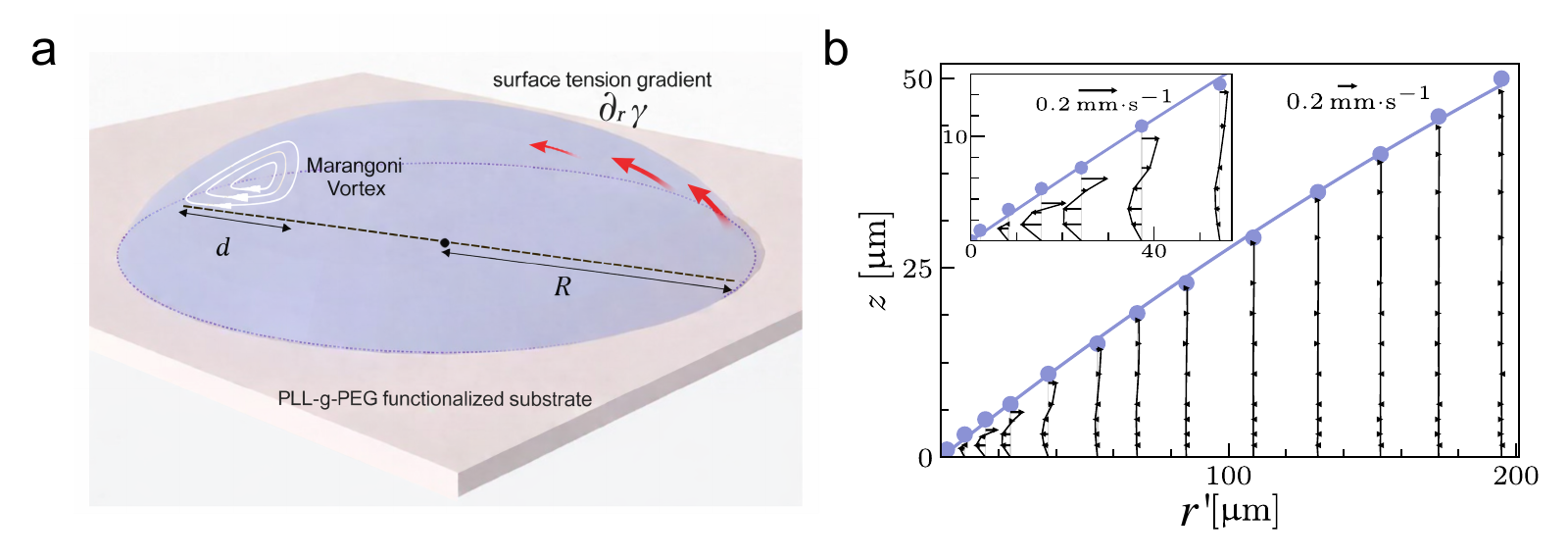}
	\end{center}
	\caption{ a) Schematic representation of an evaporating droplet of radius \textit{R} and PEG-induced Marangoni flow on a PLL-g-PEG functionalised surface. b) Measurements of the internal flow velocity of the droplet obtained with high resolution micro-PIV. Cross-sectional view of the droplet composed of M2B and PEG with molecular weight of 6 kDa: free interface (blue circles and fitting line), radial velocity (black arrows) and velocity profiles (black lines) at a distance from the contact line \textit{r}$^\prime$ = \textit{R} - \textbf{r}, where $R$ is the radius of the droplet and $r$ is the radial position in a cylindrical coordinate system with the origin at the center of the droplet. Inset: close-up of the Marangoni flow near the contact line at t = 30 s. See SI for more details.}
	\label{fig:fig3}
\end{figure}

%\subsubsection*{Marangoni flow and development of the radial corona of microtubules}
PEG functions as a depletant \cite{asakura1958interaction} and plays a pivotal role in the dynamics of microtubule and kinesin motor suspensions, as explained above. To test the role of PEG in the formation of the self-organized radial corona structure, we repeated the experiments without PEG in the active mixture.  In Figure~\ref{fig:fig2}, we compare the dynamics of the active droplet in the presence of PEG, shown in Figure~\ref{fig:fig2}(a)–(d), with the results obtained after removing the PEG (Figure~\ref{fig:fig2}(e)–(f)).
Without PEG, the active network inside the evaporating droplet exhibited a behavior different from the very beginning of the experiment. 
We did not observe the radial corona region (zone I), nor its evolution over time (See Movie S2). 
This difference can be appreciated by comparing Figures~\ref{fig:fig2}(a) and~\ref{fig:fig2}(e).
We hypothetised that the radial corona is induced by the PEG generating a Marangoni flow that leads to the alignment of the microtubule along the radial direction (Figure \ref{fig:fig3}(a)).  
Specifically, evaporation is strongest near the edge of the droplet, where the droplet also becomes very shallow. This change generates capillary forces that drive the accumulation of PEG at the contact line. This phenomenon is reminiscent of the coffee-ring effect~\cite{deegan1997}. The accumulation of PEG at the contact line establishes a PEG concentration gradient along the radial direction, which in turn generates a surface tension gradient (Figure~\ref{fig:fig3}(a)). The surface tension gradient drives a Marangoni flow that is on average directed toward the center of the droplet. The consequence of this Marangoni vortex is a zone with strong shear forces which coincides with the region where radially oriented bundles (zone I) were observed in the active network inside the droplet. 
We verified such PEG-induced Marangoni flows inside the droplet by analysing the evaporation of a droplet composed of buffer and PEG, but without the active filaments (Figure \ref{fig:fig3}(b)).
Z-resolved micro-PIV measurements showed typical Marangoni-vortex, i.e. a Marangoni-driven inward flow near the surface, compensated by a capillarity-driven outward flow near the substrate, similar to what is known from Marangoni contraction
~\cite{cira2015vapour,ramirez2022taylor,baumgartner2022marangoni}. The vortex is contained within a distance \textit{d} $\sim\SI{50}{\micro\meter}$ from the contact line (Figure \ref{fig:fig3}(b)).
In our experiments with contracting active network (Figures \ref{fig:fig1}(b) and \ref{fig:fig2}(a)), this zone was measured to have similar width \textit{d} varying between $\SI{45}{\micro\meter}$ and $\SI{80} {\micro\meter}$, with the variation likely attributable to the slightly different size of the droplets deposited on the glass substrate (volume range: $\SI{0.4} - \SI{0.6} {\micro\liter}$).   

\begin{figure}[ht!]
	\begin{center}
		\includegraphics[width=1\textwidth]{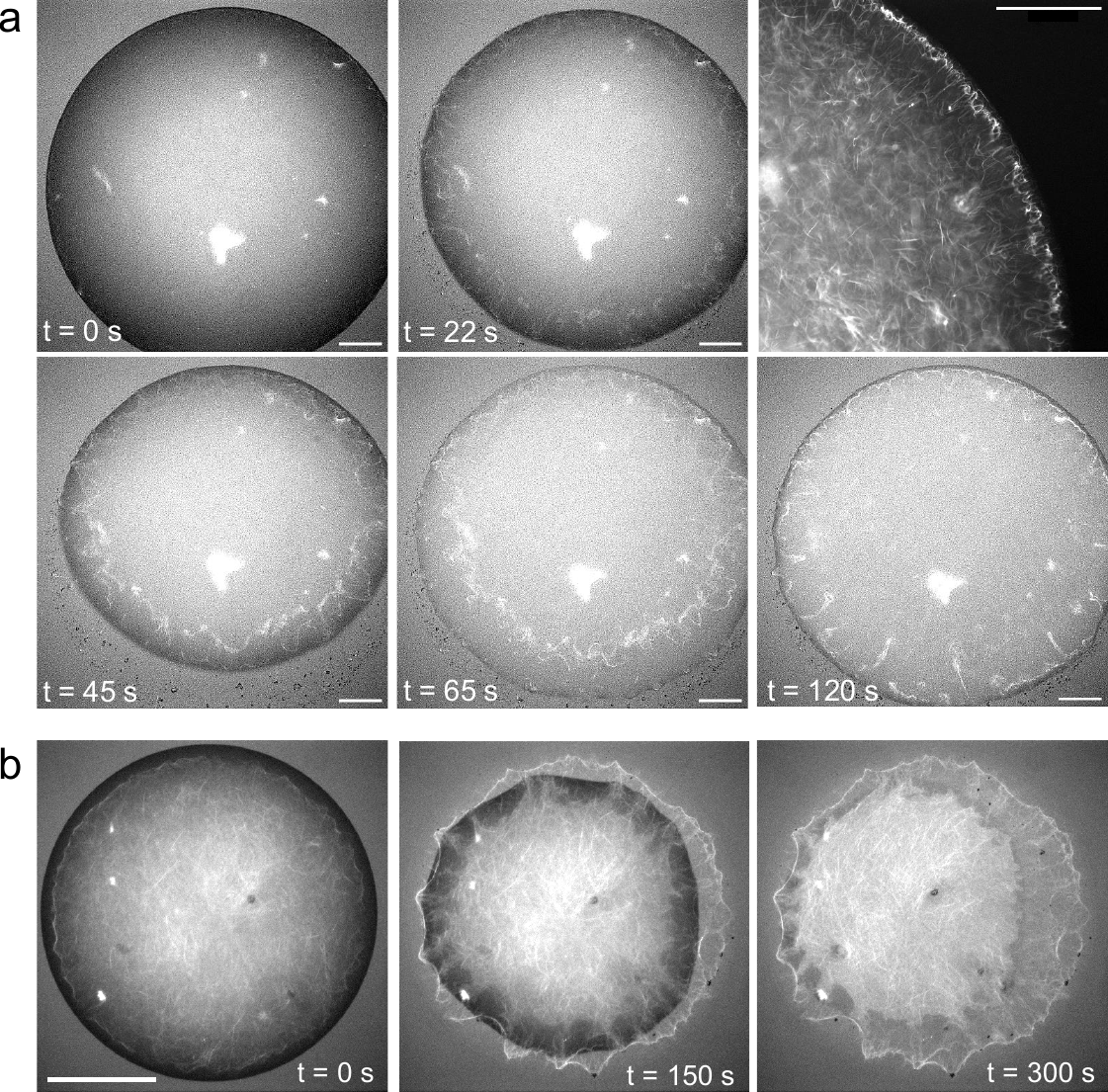}
	\end{center}
	\caption{a) Temporal evolution of a microtubule network cross-linked by kinesin motors in a solution without ATP t = 0 -- 120 s. At t = 22 s the annular region close to the contact line and rich of thin radially organized bundles of microtubules is visible (see the zoom-in view of the annular zone on the right panel). In the time frame t = 45 -- 120 s non-monotonic behaviour can be observed. Scale bar: 200 $\mu$m.  b) Micrographs of the evaporating droplet containing microtubule network with PEG in the absence of motor proteins. Scale bar: 500 $\mu$m. }
	\label{fig:fig4}
\end{figure}

We further characterised the formation of the radial corona by investigating the contribution of the motor proteins in cross-linking microtubules compared to the role of Marangoni flow in forming the radial alignment of filament bundles. For this purpose, we performed experiments with microtubules and motor-proteins similar to the setup in Figure 1(b), but without ATP (Figure~\ref{fig:fig4}(a) and Movie S3). As kinesin-1 still binds to microtubules in the absence of ATP \cite{vugmeyster1998}, a cross-linked network can also form under these experimental conditions. We compared the outcomes with those of experiments in which both motor proteins and ATP were not included in the sample suspension (Figure~\ref{fig:fig4}(b) and Movie S4). Interestingly, in both cases we observed the formation of a radial corona of aligned microtubules in zone I similar to the one shown in Figure~\ref{fig:fig1}(b), revealing that neither the activity of the motors nor their mere presence is required for the radial corona to form. We conclude that Marangoni flow plays the main role for the arrangement of microtubule bundles in zone I and the formation of the radial corona of the active network in Figure~\ref{fig:fig1}(b). However, in the system without motor proteins added to the sample suspension, we observed a distinctive droplet-microtubule behaviour that we describe in details in the next section.
\\
\\
\textbf{Droplet dynamics without motors}
\\
The removal of motor proteins from the sample led to distinct spatiotemporal dynamics during droplet evaporation. At the beginning of the experiment (Figure~\ref{fig:fig4}(b), $t = 0$ s), the microtubule bundles were randomly distributed filling the droplet. 
The droplet began to evaporate and gradually shrank over time (Figure~\ref{fig:fig4}(b), $t = 150 $ s). Unlike the case with motors, the network and the droplet did not exhibit any synergistic behavior. The droplet footprint shrank via a stick–slip mechanism, moving inward relative to the microtubule network. The entangled network, which was aggregated by the shrinking droplet compartment and is visible in Figure~\ref{fig:fig4}(b) at $t = 300~\mathrm{s}$ as a region of high fluorescence intensity, partially adhered mechanically to the substrate. In addition, microtubule bundles formed a radial arrangement (Figure~\ref{fig:fig4}(b), $t = 300~\mathrm{s}$). 
Comparing this behavior with that of a system containing motors in the absence of ATP shows that the cross-linking ability of the motor proteins plays a crucial role in forming a network capable to overcome substrate adhesion.\\
We also tested the effect of the size of microtubule bundles on the aggregation behavior of the network in the absence of motors by using PEG with different molecular weights.
Figure S3 in the SI shows the networks generated using PEG with molecular weights of 6 kDa and 100 kDa, which can be compared to the network formed with PEG of 20 kDa. Shorter PEG chains produce thinner bundles and a less interconnected network, whereas longer PEG chains generate thicker bundles and greater connectivity. This trend is consistent with the fact that depletion attraction scales with molecular weight\cite{ZASADZINSKI20051621}.
Additionally, increasing the PEG molecular weight influenced both the aggregation in zone II and the distribution of microtubule bundles in zone I adhering to the substrate. However, it did not prevent the filaments from adhering to the substrate.

\subsection*{Dynamics of the contact line}
Figure~\ref{fig:fig1}(b) clearly shows that the droplet and the embedded gel-like active network evolve with distinctive dynamics. In addition to the radial corona pattern discussed in detail above, we observed that the radii of both the contact line and the isotropic network in zone II initially shrank and then spread (see Figure~\ref{fig:fig1}(c)), resembling a dewetting-wetting process of the substrate. Throughout all stages of this dynamic process, the network moved relative to the footprint, consistently in the same direction. 
Particularly, during the first $\sim 10-15$~seconds, the droplet footprint diameter shrank by about $\sim 24$\%. Afterwards, until $t\sim 30$~s, the droplet spread again to a larger footprint diameter up to $\sim 90$\% of its initial size (Figure \ref{fig:fig1}(c)).  
During these later stages, the isotropic network (zone II) expanded through the entire droplet (Figure \ref{fig:fig1}(b) at t = 30~s). The faster expansion of the central network (zone II) compared to the contact line was evident also by considering their spreading velocity (see Figure S4 in the SI). Shrinking-spreading behaviour was observed in every setup containing motor proteins, including control experiments where either PEG or ATP were removed from the sample suspensions. 
However, the control experiments showed that the non-monotonic contact line expansion occurs on a distinct timescales compared to the case that included both ATP and PEG. These analyses helped us verify the hypothesised driving mechanism as explained below.
\subsubsection*{Gel-like network and cross-linking by the motor proteins
}\label{glikestr}
When we removed PEG from the experimental sample, we did not observe the radial distribution of bundles and pattern formation of the active network inside the droplet, nor its evolution over time.
\begin{figure}[h!]
	\begin{center}
		\includegraphics[width=0.6\textwidth]{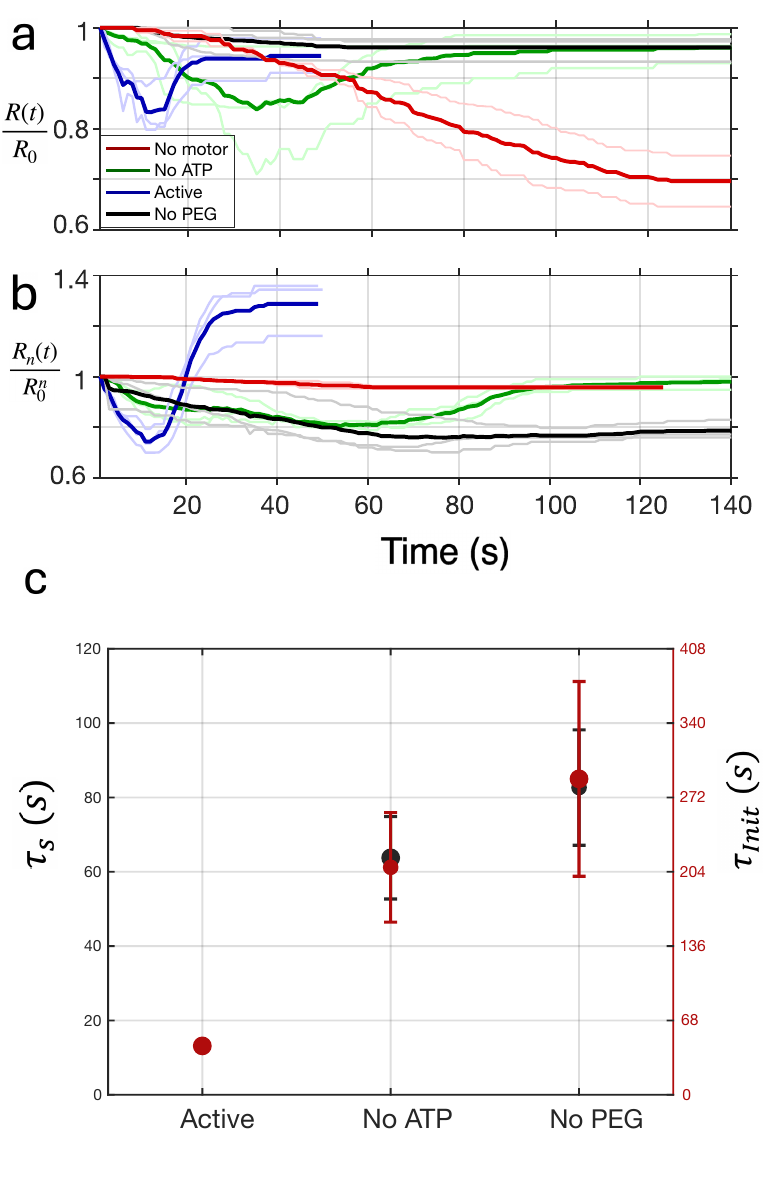}
	\end{center}
	\caption{Dynamics of the contact line (a) and the network radius (b) under different experimental setups. %Red: No motors; Green: No ATP; Black: No PEG; Blue: Active droplet. 
    Dark colors represent the averages, and lighter colors individual experiments. The radius is normalized by the initial radius. In all cases with motors, the network radius first decreases and then increases. This behavior is less evident in the average for the No PEG case while is clear in individual experiments. The reduced evidence in the average arises because the non-monotonic behavior occurs at different times in different experiments (see Figure S6 in SI). In (a) and (b), $R_0$ ($R_0^{n}$) denotes the initial radius of the droplet (network), and $R(t)$ ($R_n(t)$) denotes the radius of the droplet (network) over time. (c) The average turnover timescale, $\tau_s$ (black), and the initial retraction timescale, $\tau_{Init}$ (red), are shown for different conditions. The initial network retraction timescale, $\tau_{Init}$, defined as the timescale over which the network radius contracts to $90\%$ of its initial value, scales linearly with the turnover timescale, $\tau_s$. Specifically, the timescale over which the network starts expanding, $\tau_s$, scales linearly with $\tau_{Init}$, with a proportionality factor of approximately $3.4$. Error bars represent the standard deviation across different experiments.
 }
	\label{fig:radiusdy}
\end{figure}

We still observed the presence of a microtubule network filling the droplet with thinner bundles (Figure~\ref{fig:fig2}(e) at $t = 0$ s). During the evaporation, this network continuously moved toward the center of the droplet, resembling the zone II we described previously. A slightly receding contact line motion (See Movie S2) could be observed (Figure~\ref{fig:fig2}(e) at $t = 30 - 50$ s) while the radial corona (zone I) was absent. In zone II the aggregation of the active network was not as pronounced as in droplets containing PEG (Figure~\ref{fig:fig2}(a) vs Figure~\ref{fig:fig2}(e)). We attribute this difference to the absence of PEG acting as a depletion agent that leads to formation of bundles. In this setup, the microtubule solution did not undergo phase separation into bundles, and the resulting network was less interconnected than in the presence of PEG. However, we still observed a contraction of the network as shown in Figure~\ref{fig:fig2}(f).
The formation of the central gel-like structure in zone II in our experiments likely arose from the presence of the motor proteins and their capacity to cross-link the network albeit less efficiently than in the presence of PEG. Indeed, motor proteins can cross-link filaments in a network even without PEG. However, their interconnection is not as pronounced as in the case with PEG since the osmotic pressure exerted by the depletant is missing. After around $\sim 70 $ s a slight spreading involving the network inside the droplet can be detected also in this case, but it does not occur at the contact line (See Movie S2 in SI). It is evident that the shrinking–spreading process was not driven by the presence of PEG (and the strong Marangoni flow it induced), but rather by the action of motor proteins, as we hypothesized.
%%%
To confirm our hypothesis and further elucidate the role of motor proteins in this process, we analysed the shrinking–spreading behavior of the experimental system including PEG but in the absence of ATP, both with and without cross-linking motor proteins (Figure~\ref{fig:fig4}(a) and Figure~\ref{fig:fig4}(b), respectively).
In the presence of motor proteins we observed the contact line and the network shrinking and forming the central zone II when the droplet started evaporating. During the evaporation time  the shrinkage was followed by the spreading of both zone II and contact line similar to the case shown in Figure~\ref{fig:fig1}(b). 
 This non-monotonic dynamics disappeared in the system without motor proteins (Figure~\ref{fig:fig4}(b)), in which the contact line of the droplet receded over the evaporation time, leaving behind the network adhering on the substrate. This confirmed our hypothesis about the main role of the motor proteins in the dynamics of the contact line.
 When comparing the experiments including motor proteins, an interesting difference between the experiments with and without ATP lies in the timescale at which the non-monotonic behavior is observed, with the system lacking ATP shrinking for a longer period before spreading.

We summarize the dynamics of the systems resulting from the different setups in the graphs in Figure \ref{fig:radiusdy}. 
These experiments verified our hypothesis that the observed cycle of contact line contraction and expansion arises from passive crosslinking interactions between microtubules and kinesin motors. Notably, the graph shows that when motors are omitted from the experimental solution, the droplet radius decreases monotonically over time. Furthermore, the presence of ATP and the resulting extensile activity significantly accelerate the turnover dynamics, reducing the characteristic timescale associated to the non-monotonic behaviour by an order of magnitude.
Across all experiments with different initial conditions, we observed that a sharper initial decay in the contact line radius correlates with an earlier onset of contact line expansion (see Figure \ref{fig:radiusdy} (c)).
We conjectured that the spreading process was possibly due to the altered concentration of the constituents in the salty buffer caused by the water loss by evaporation and its effect on motor proteins. Specifically, studies showed that as ionic strength in solution increases, the overall efficiency of the kinesin cross-linking reaction decreases \cite{TUCKER1997,thorn2000}. This likely destabilised the network in zone II, and ultimately led to droplet expansion.
The ionic strength of the M2B buffer used during the experiments could be determined by using the Debye-Hückel equation and in our case it was around 60 mM. Using a drop shape analyzer, we measured the volume of droplets deposited on the substrate at approximately 40 s (Figure S5 in the SI), corresponding to the turnover time in the system without ATP (Figure \ref{fig:fig4}(a)). By this time, the reduction in volume led to a strong increase in ionic strength, which can be estimated to reach approximately 100 mM.
\begin{figure}[ht!]
	\begin{center}
		\includegraphics[width=0.7\textwidth]{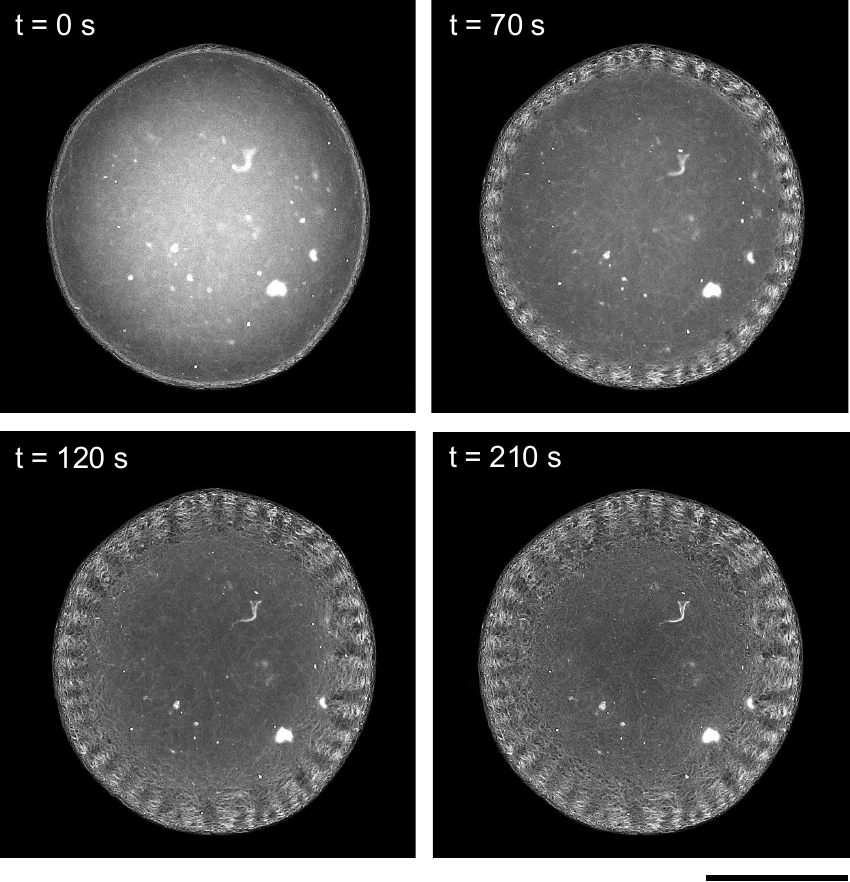}
	\end{center}
	\caption{a) Temporal evolution of a microtubule network in a droplet without ATP and containing motor proteins, PEG and addition of $\text{MgCl}_{2}$, bringing the ionic strength to 100 mM. The formation of the radial corona can be observed over time. However, no formation of central network (zone II) is visible. Scale bar: 500 $\mu$m. }
	\label{fig:fig6}
\end{figure}
We tested our conjecture by repeating the evaporating droplet experiment with PEG and motor proteins in the absence of ATP and increasing the M2B ionic strength to 100 mM with the addition of $\text{MgCl}_{2}$ (Figure \ref{fig:fig6} and Movie S5). Under these experimental conditions the network showed the generation of radial corona over the evaporation time, confirming the Marangoni flow as main driving process in the generation of the radial organization. However, no zone II formation was visible proving the influence of the increased salt concentration on the network behaviour. 
Our interpretation about the role of salt in the functionality of the motor proteins is supported by the observation that the turnover time decreases dramatically in the active system (network with motor proteins fueled by ATP in Figure \ref{fig:fig1}). This can be explained by the presence of active inward radial flows generated by active extensile flows from the radial orientation of microtubules combined to the Marangoni flow, which induced the advection of salt into the central gel-like region of the droplet, thereby accelerating the dewetting-to-wetting transition (See the next section for a more detailed analysis).
\\
 
 \subsubsection*{Lubrication theory}
We model the evaporating active droplet as a thin fluid layer of height $h(r,t)$ and lateral extent $R$, with $h \ll R$. This separation of scales justifies a lubrication description in which vertical momentum balance is dominated by viscous stresses, while horizontal flow is driven by gradients in pressure, surface tension, and internally generated active stress.

Evaporation is spatially nonuniform, being strongest near the contact line. As in passive evaporating droplets, this induces a compensating radial flow within the droplet interior. In classical colloidal systems, this flow is governed by a balance between capillary pressure gradients and viscous resistance, leading to either contact-line recession or material accumulation depending on the relative magnitude of evaporation and flow rates~\cite{Kaplan1,Kaplan2}. This asymptotic structure has been developed in detail for passive droplets and particle-laden films and forms the baseline for our analysis. In passive colloidal droplets, lubrication models have shown that evaporation-induced flow is controlled by a balance between viscous resistance and capillary or Marangoni stresses, with deposition patterns selected by a small number of dimensionless groups. More elaborate multiphase formulations further demonstrate how internal microstructure can feed back on the flow through a transition from Stokes to Darcy-like behavior as material densifies near a deposition front. 

The present system follows the same asymptotic logic but differs qualitatively in one crucial respect: the microtubule–kinesin network generates stresses internally and dynamically. Activity therefore plays a role analogous to a time-dependent, internally regulated rheological field, rather than a passive resistance or imposed boundary condition \cite{nejadarx}.

To explain the origin of the radial corona shown in Figure \ref{fig:fig1}, we note that evaporation is strongest near the contact line, producing a narrow annular region of strong shear. In this region, microtubule bundles experience sustained extensional flow and align along the principal stretching direction. Because the shear is predominantly radial, the nematic director $\textbf{m}$ aligns radially, $\mathbf m \approx \hat{\mathbf r},$ forming the observed corona. This alignment does not require motor activity: it is set by evaporation-driven Marangoni flow and geometrical confinement. Activity becomes important only once alignment has occurred and nematic order paramter becomes non-zero ($S\neq0$). 
We note that because the droplet remains thin, the microtubule network may be described as an effectively two-dimensional active nematic confined to the plane of the substrate. We therefore characterize the internal organization of the network by the unit director field ${\bf m}(x,y,t)$, representing the local in-plane alignment of microtubule bundles. Then, standard lubrication-type arguments yield an evolution equation for the film thickness $h(x,y,t)$ that evolves according to the following nonlinear PDE
$$\partial_t h + \nabla\!\cdot(h\mathbf u) = -J(x,y),$$ where ${\bf u}(x,y,t)$ is the depth-averaged in-plane velocity and $J(x,y,t)$ is the evaporation rate. 
In the lubrication limit, the in-plane flow is linked to gradients of stress integrated across the film thickness. Retaining only leading-order contributions, the depth-averaged velocity may be written as (see SI for details)
%\begin{equation}
%\mathbf u \;\sim\;-\frac{H^2}{\mu}\,\nabla %p\;+\;\frac{H}{\mu}\,\nabla\gamma\;+\;\frac{H^2}%{\mu}\,\nabla\!\cdot\!\big(\zeta\,\mathbf n\mathbf %n\big).\label{velocityav}
%\end{equation}

\begin{equation}
\langle \tilde{v}_r\rangle  = \frac{\tilde{h} \partial_{\tilde{r}} \tilde{\gamma}}{2\mu} -\frac{\tilde{\zeta} \tilde{h}^2 S}{3 \tilde{r}\mu}.\label{velocityav}
\end{equation}
where the parameters with tilde are rescaled with the aspect ratio of the droplet $\epsilon = h/R$, $ \tilde{\zeta}> 0$ denotes
extensile activity characteristic of microtubule–motor suspensions, and $\mu$ is the viscosity of the fluid.
Here, the two terms correspond respectively to: (i) Marangoni flows, generated by evaporation-induced PEG concentration gradients, which drive an inward surface flow and promote droplet contraction, and (ii) active extensile flows, generated by the microtubule–kinesin network. 
 
The average velocity in Eq. (1) suggests that
both extensile activity and the surface tension gradient generated by PEG (with $\partial_{\tilde{r}} \tilde{\gamma} < 0$ as the PEG concentration decreases toward the center) drive, on average, a fluid flow toward the center of the droplet and lead to contraction of the network and the contact line at short times. This also supports our hypothesis that extensile activity and Marangoni flow advect salt toward the center of the droplet. Over longer timescales, the advection of salt toward the center of the droplet (where the cross-linked network resides) likely impairs the ability of the motors to maintain network cross-linking, ultimately leading to network dissociation and expansion (see Figure~\ref{fig:radiusdy}(c)). Assuming that the effect of salt on network unbinding is rapid compared to the evaporation timescale, this implies that the timescale for the initial decay of the contact line and network radius is proportional to the turnover timescale. Indeed, in the experiments the timescale for the initial contraction of the network scales linearly with the turnover timescale (see Figure~\ref{fig:radiusdy}(c)).
Eq. \ref{velocityav} introduces a Marangoni timescale $\tau_M\sim \mu \tilde{R}^2/(\tilde{h} |\Delta \tilde{\gamma}|)$, where $|\Delta \tilde{\gamma}|$ is the change in the surface tension from the contactline to the center of the droplet, an active timescale $\tau_a\sim \mu \tilde{R}^2/(\tilde{\zeta} \tilde{h}^2)$. The evaporaton timescale $\tau_e\sim\tilde{h}/\tilde{J}$, is
the smallest timescale in our experiments according to the measurements without PEG (see Figure \ref{fig:radiusdy}(c)).
In our experiments, PEG gradients and the consequent Marangoni flows are already formed at the onset of the experiment and this implies that the timescale of the capillary flow  is smaller than the Marangoni timescale of order $5\,\text{s}$  an upper estimate for the onset of recording. The ratio of the active stresses to the Marangoni stresses in physical units is given by $\mathcal A = \tau_a/\tau_M$. Our measurements show that the surface tension variation due to PEG in the droplet $\Delta \gamma$ is in the interval $(1 -10)\,\text{mN/m}$ (See Table S1 in SM), and the aspect ratio of the droplet is estimated as  \(\epsilon= H/R \sim 0.1\). The ratio of the active timescale to the Marangoni timescale in Figure \ref{fig:radiusdy}(c), which is $ \mathcal A \sim 0.2$, then gives an estimation for the activity $\zeta \sim (0.05-0.5)\,\text{Pa}$. Considering an active length of $1-10 \mu m$, this has a reasonable overlap with the measured activity in a similar setup $\zeta \sim (0.02-0.2)\,\text{Pa}$~\cite{velez2024probing,strübing2020wrinkling}.

Since the same flows that contract the droplet advect the salt towards the network at the center, the theory suggests that the timescale of the motor dysfunction, which originates from salt advection, should be proportional to the retraction timescale. As shown in Figure~\ref{fig:radiusdy}(c), the initial retraction time $\tau_{Init}$ is linearly related to the turnover time $\tau_S$.

\section*{Conclusion}
We studied the dynamics of an active suspension of microtubules and kinesin motors inside an evaporating droplet on a solid surface at room temperature. PEG, the depletion agent used to bundle the microtubules, induced Marangoni flows that led to pattern formation, resulting in a radially aligned arrangement of microtubules at a distance from the center.
We observed that contact line first contracts and then expands over time.
Our control experiments verified that the observed cycle of contact line contraction and expansion arises from passive crosslinking of microtubules by kinesin motors. Indeed, when motors are omitted from the solution, the droplet radius decreases monotonically over time at a rate comparable to that driven by evaporation alone. Furthermore, the presence of ATP and the resulting extensile activity significantly accelerates the turnover dynamics, reducing the characteristic timescale by an order of magnitude.

Overall, our results show that at short times, motor binding to the filaments forms a dense network of microtubules at the center of the droplet, which becomes progressively denser and leads to contact line shrinkage. Additionally, the self-organized structure of radially aligned microtubules formed by Marangoni flow drives active extensile flows toward the center, further contributing to contact line contraction. Similarly, the average Marangoni flow is directed toward the center, reducing the contact line radius.
At longer times, we showed that it is plausible to hypothesize that the combined effects of active and Marangoni flows could advect salt toward the center, where the dense network resides. The resulting high ionic concentration likely disrupts kinesin motor binding ability—and thus filament cross-linking—leading to network disassembly and subsequent expansion of both the network and the contact line.
Across all experiments with different experimental set up, the timescale for the initial contraction of the network is proportional to the timescale of onset of network expansion.
Also this behavior arises from the dual effect of inward-directed flows, which cause the contact line to contract and to concentrate salt at the center promoting motor unbinding with subsequent network and contact line expansion. By using the combination of experiments and a lubrication theory, we show that active extensile flows and Marangoni flow reduce the droplet shrinkage timescale and the timescale over which expansion happens.
The droplet thus acts as a self-organized active Marangoni system, whereby evaporation creates shear which aligns the nematic locally, and thus promotes extensile active stresses that speed up the dynamics. Crucially, the director field plays the role of an internal geometric switch: it converts isotropic activity into directed forces only when alignment is present. 

In this study, we investigated evaporating droplets containing biological filamentous material and analyzed their distinctive behavior. In future work, we will focus on understanding the implications of the self-organized patterns of active filaments observed here, particularly in contexts such as the origin of complexity in living matter.
\\
\\
\\

\section*{Materials and Methods}\label{methodsection}
\subsubsection*{Polymerization of Microtubules}
Microtubules were polymerized from 2.7 mg/ml HiLyte labeled
porcine brain tubulin (Cytoskeleton, Inc., U.S.A.) in M2B with 5 mM MgCl$_{2}$, 1 mM GTP, and 5\% DMSO
at 37$^{\circ}$C for 30 min. The microtubules were stabilized with 7 $\mu$M taxol and mixed with 0.5 mg/ml glucose, 0.65 mM dithiothreitol (DTT), 0.2 mg/ml glucose oxidase (Sigma G2133), 0.05 mg/ml catalase (Sigma
C40) and 2.4 mM Trolox (Sigma 238813) to avoid photobleaching. Depending of the specific case, 1\% PEG with molecular weight 6, 20 and 100 kDa was added. For the experiments on active droplets kinesin 401 was added. The plasmid that codes biotin-labeled kinesin 401 (K401) was a gift from Jeff Gelles (pWC2 - Addgene plasmid \# 15960; http://n2t.net/addgene:15960; RRID Addgene 15960)\cite{Subramanian445}. Kinesin 401 was purified as previously published \cite{Gilbert,Young} and the kinesin-streptavidin complexes were prepared by mixing 0.2 mg/mL kinesin 401, 0.9 mM DTT, and 0.1 mg/ml streptavidin (Invitrogen, S-888) dissolved in M2B and
incubated on ice for 15 min. 4 $\mu$l of this mixture was mixed with ATP at a final concentration of 1 mM, 1.7 $\mu$l of pyruvate kinase/lactic dehydrogenase (PK/LDH, Sigma, P-0294), 32 mM phosphoenol pyruvate (PEP, VWR AAB20358-06) to form a solution of active clusters that was added to the microtubules solution described above.  The microtubules and the active clusters solution were mixed 15 min before the onset of the experiments. For the control experiments where individual components were removed form the solution, the equivalent amount of volume was replaced by M2B. For the experiments with increased ionic strength, a suitable volume of 1M $\text{MgCl}_{2}$ was added to the solution 15 min before starting the experiment to reach the desired final ionic strength. 

\subsection*{Preparation of functionalised Glass Surfaces}\label{moredet}
Microscope coverslips were cleaned by washing with 100\% ethanol and rinsing with deionized water. They were then sonicated in acetone for 30 minutes and incubated in ethanol for 10 minutes at room temperature. This was followed by incubation in a 2\% Hellmanex III solution (Hellma Analytics) for 2 hours, extensive rinsing with deionized water, and drying with filtered airflow.
When substrate functionalisation was required, the cleaned coverslips were immediately activated in oxygen plasma (FEMTO, Diener Electronics, Germany) for 30 seconds at 0.5 mbar, and subsequently incubated in 0.1~mg/mL poly(\textit{l}-lysine)-graft-poly(ethylene glycol) (PLL-g-PEG) (SuSoS AG, Switzerland) in 10~mM HEPES buffer at pH~7.4, at room temperature for 1 hour on Parafilm (Pechiney, U.S.A.). Finally, the coverslips were slowly lifted off, and the remaining PLL-g-PEG solution was removed to achieve complete surface dewetting.

\subsection*{Image Acquisition and Analysis}
Image acquisition was performed using an inverted fluorescence microscope (Olympus IX-71) with 4$\times$, 10$\times$, or 20$\times$  objectives (Olympus, Japan), depending on the experimental setup. For excitation, a Lumen 200 metal arc lamp (Prior Scientific Instruments, U.S.A.) was used. Images were recorded with a CCD camera (CoolSnap HQ2, Photometrics). Frames were acquired at 1~Hz. Image analysis was performed using ImageJ software to measure the shrinkage and spreading of the droplets.

\bibliography{bibliography}
\bibliographystyle{naturemag}

\setcounter{equation}{0} \setcounter{figure}{0}
\renewcommand{\theequation}{S\arabic{equation}}
\renewcommand{\thefigure}{S\arabic{figure}}
\clearpage

\appendix

\section*{Supplementary Methods}

\subsection*{PIV measurements}
The internal flows of evaporating droplets, mixtures of M2B buffer and PEG, were quantified using micro particle image velocimetry (micro-PIV).
Three different molecular weights of PEG were used (6, 20, and 100 kDa). The experiments were done inside a humidity control chamber (a 10 cm cubic chamber), at room temperature. The humidity was set by continuously injecting a predefined mixture of dry nitrogen and nitrogen saturated with water vapor behind gas-permeable membranes at the side-walls of the chamber. Droplets with initial volumes of 1 $\mu$L were placed on substrates (functionalized glass coverslides, $24\times 24$ mm).
Polystyrene microspheres (Thermo Fisher Scientific F8809, diameter of 200 nm) were used as flow tracers, with a mass fraction of $7.8 \times {10^{-5}}$ of the particle stock solution in the mixtures. The particle concentration was 
$ 1.7 \times {10^{10}}$ particles/mL. 
The particles within the drops were observed with an inverted epifluorescence microscope (Nikon Eclipse Ti2) for the PIV measurements. The microscope was equipped with a water immersion objective (Nikon CFI APO LWD 20-8192  WI) for diffraction-limited imaging, with a numerical aperture of 0.95. Images of the particles were captured with a high-speed camera (Phantom VEO 4K 990L, imaging speed at 900 fps), quickly switching between planes parallel to the substrate by automating the focus system of the microscope. Once the drop was deposited on the substrate, the surface of the substrate was visible under the microscope by focusing to the particles located at the contact line. After finding this reference plane the recording starts. Thus, the time between the deposition of the droplet and the start of the micro-PIV recording is not defined. A side-view camera (Point Grey Grasshopper2, imaging speed at 27 fps) mounted on the microscope simultaneously recorded the deposition of the drop, the macroscopic apparent shape of the droplet, and the start of the micro-PIV recording. The later was detected when the epifluorescence excitation light was observed, after opening the shutter synchronously, due to scattering.  The side-view camera was equipped with a telecentric macro lens (Thorlabs Bi-Telecentric lens 10$\times$8192 , working distance 62.2 mm).

\subsection*{PIV measurements analysis}

The recorded images of evaporating droplet for the PIV measurements were analyzed with an in-house Python developed code to quantify the internal flows. The data was evaluated through a cross-correlation with correlation-averaging over 150 frames \cite{meinhart2000piv}. An adaptive interrogation window size method is included in the algorithm. First, single-pixel correlations were calculated for the entire image and all displacements within a predefined search range\cite{westerweel2004single}. Instead of correlating intensity values directly, we used the dot product of the gradient (first order differences). Then, the correlations were integrated over interrogation windows of various sizes by convolution with a square kernel of the desired size. 
We used five different interrogation window sizes of 4, 8, 16, 32, and 64 px side-length. 
The final correlation maps were then evaluated by a weighted average between the different window sizes, using the mean square of the intensity gradient values in the interrogation window and a size-dependent bias as weight. 
The method was implemented through the Python API of Tensor Flow. 
From this method, 2D velocity fields were obtained for different $z$-planes parallel to the substrate.
The radial component in each $z$-plane was calculated by azimuthally averaging over $100~\mu$m. The corresponding experimental uncertainties of the radial velocities were estimated using the standard deviation over the weighted interrogation windows. With the radial velocities we reconstructed the velocity profiles for different experimental conditions. Close to the free surface, correlations were picked up preferentially from below the focal plane because no particles were outside the drop. This leads to a shift in the correlation plane relative to the focal plane. Accordingly, we applied a correction to the $z$ location of the velocity signal relative to the distance from the free surface.
To obtain this correction, we estimated the point spread function and the correlation sensitivity as a function of distance to the focal plane, and convolved this sensitivity with a unit-step function in $z$ for the particle density. For these measurements, we modeled the depth of correlation as a Gaussian with $1.5~\mu$m standard deviation.

\subsection*{Surface tension measurements of different buffer mixture}
The surface tension measurements were conducted by the pendant drop method \cite{hansen1991surface}. Briefly,  a small drop of liquid was suspended at the end of a needle to form a tear-shaped pendant drop. The drop shape was recorded, and the optical image fitted with a model based on the Young-Laplace equation, to determine the surface tension. For each solution, the surface tension of 8 drops of 2.5 $\mu$L was measured (in room conditions, T = 20 $^{\circ}$C, RH = 45\%, Data Physics OCA 20). Ten images were collected for each drop in 1 s of recording time. The surface tensions $\gamma$ were calculated as an average of these measurements with an average error.

\begin{table}[h!]
	\centering
	\begin{tabularx}{0.8\textwidth}{
			| >{\raggedright\arraybackslash}X
			| >{\centering\arraybackslash}X |
		}
		\hline
		Composition & Surface tension $\gamma$ ($\mathrm{mN\,m^{-1}}$) \\
		\hline
		Water & $72.0 \pm 0.4$ \\
		\hline
		PEG 5\% & $62.2 \pm 0.1$ \\
		\hline
		PEG 1\% & $62.9 \pm 0.2$ \\
		\hline
		PEG 0.5\% & $63.1 \pm 0.2$ \\
		\hline
		M2B 5\% & $73.3 \pm 0.2$ \\
		\hline
		M2B 1\% & $72.6 \pm 0.3$ \\
		\hline
		M2B 2.5\% + PEG 2.5\% & $61.4 \pm 0.2$ \\
		\hline
		M2B 2.5\% + PEG 0.25\% & $61.8 \pm 0.1$ \\
		\hline
	\end{tabularx}
	\caption{Surface tension measurements of M2B buffer plus PEG.}
	\label{table:1}
\end{table}

\subsection*{Lubrication Theory}
\label{lubricationtheor}
In this section, we apply the lubrication approximation to understand the effects of the self-organized aster structure formed by active microtubules, as well as the effect of the surface tension gradient created by the PEG gradient along the radial direction, on the flows generated in the droplet. We focus on the bulk flows and the region far from the contact line, so we do not consider a precursor wetting film and the boundary condition associated with that.
The unit vector normal to the droplet interface $\mathbf{n}$, the unit tangent vectors along the interface $\mathbf{t}$, and the droplet curvature $\kappa$ on the interface are defined as follows:
\begin{align}
	&\mathbf{n}=\frac{\left(-\partial h / \partial r, 1\right)}{\sqrt{\left(\partial h / \partial r\right)^2+1}},\\
	&\mathbf{t}=\frac{\left(1, \partial h / \partial r\right)}{\sqrt{\left(\partial h / \partial r\right)^2+1}},\\
	&\kappa=\nabla \cdot\left[\left(1+\left|\nabla h\right|^2\right)^{-1 / 2} \nabla h\right],
\end{align}
where \( h(r,t) \) denotes the droplet thickness.  
To apply the lubrication approximation,  
we assume that the droplet thickness \( h \) is much smaller than the droplet radius \( R  \),  
and define the small dimensionless parameter \( \epsilon = \frac{h}{R} \ll 1\).  
We then expand the fields and parameters in terms of the small dimensionless quantity \( \epsilon \):
\begin{align}
	r &= R \tilde{r},\quad z = \epsilon R \tilde{z},\quad h = \epsilon R \tilde{h}, \\
	v_r &= V \tilde{v}_r,\quad v_z = \epsilon V \tilde{v}_z,\quad t = \frac{R}{V} \tilde{t}, \\
	p &= \frac{\mu V}{\epsilon^2 R} \tilde{p},\quad J = \rho V \epsilon \tilde{J},\quad \zeta \sim \epsilon^{-2} \tilde{\zeta},
\end{align}
where the tilde variables are dimensionless. The fluid velocity in the radial direction is denoted by \( v_r \), the velocity perpendicular to the substrate by \( v_z \), and the pressure inside the fluid by \( p \), while the evaporation rate is represented by \( J \). All of these quantities are functions of the radial distance from the center of the droplet, \( r \), and the vertical distance from the substrate, \( z \).

In addition, we define the velocity scale as \( V \sim \epsilon^0 \), and introduce the activity coefficient \( \zeta \). Using these scalings, and assuming axisymmetry (i.e., no dependence on the azimuthal angle \( \theta \)), the components of the stress tensor \( \boldsymbol{\sigma} \) in cylindrical coordinates are given by:
\begin{align} &\sigma_{r r}  =-\epsilon^{-2} \tilde{p}+2 \mu \frac{\partial \tilde{v}_r}{\partial \tilde{r}} -\epsilon^{-2} \tilde{\zeta} Q_{rr},\\ &\sigma_{\theta \theta}  =-\epsilon^{-2} \tilde{p}+2 \mu\left(\frac{\tilde{v}_r}{\tilde{r}}\right) - \epsilon^{-2} \tilde{\zeta} Q_{\theta \theta}, \\ &\sigma_{z z}  =- \epsilon^{-2} \tilde{p}+2 \mu \frac{\partial \tilde{v}_z}{\partial \tilde{z}}-\epsilon^{-2} \tilde{\zeta} Q_{zz}, \\ & \sigma_{\theta z}=\sigma_{z \theta} =0, \\ & 
	\sigma_{r \theta}=\sigma_{\theta r}=-\epsilon^{-2} \tilde{\zeta}  Q_{r \theta}, \\ & 
	\sigma_{r z}=\sigma_{z r}  =\mu\left(\epsilon^{-1}\frac{\partial \tilde{v}_r}{\partial \tilde{z}}+\epsilon \frac{\partial \tilde{v}_z}{\partial \tilde{r}}\right) ,\end{align}
where $\textbf{Q}= S (\textbf{m}\textbf{m}-\textbf{I}/3)$ is the nematic tensor associated with the nematic order parameter of MTs, where $S$ shows the magnitude of the order and $\textbf{m}$ shows direction of the order, and $\zeta>0$ corresponds to extensile activity. The unit vectors perpendicular to the interface, $\mathbf{n}$, and parallel to the interface, $\mathbf{t}$, as well as the curvature of the droplet at the interface, $\kappa$, can be expanded in terms of $\epsilon$ as:
\begin{align}
	&\mathbf{n}=\frac{\left(-\partial h / \partial r, 1\right)}{\sqrt{\left(\partial h / \partial r\right)^2+1}}\sim [-\partial_{\tilde{r}} \tilde{h} \epsilon, (1-(\partial_{\tilde{r}} \tilde{h})^2 \epsilon^2/2)],\\
	&\mathbf{t}=\frac{\left(1, \partial h / \partial r\right)}{\sqrt{\left(\partial h / \partial r\right)^2+1}}\sim [(1-(\partial_{\tilde{r}} \tilde{h})^2 \epsilon^2/2),\partial_{\tilde{r}} \tilde{h} \epsilon],\\
	&\kappa= \boldsymbol{\nabla} \cdot (\textbf{n}/|\textbf{n}|)= 
	\boldsymbol{\nabla} \cdot
	[
	\frac{\left(-\partial h / \partial r, 1\right)}{\sqrt{\left(\partial h / \partial r\right)^2+1}}]=\boldsymbol{\nabla} \cdot [-\partial_{\tilde{r}} \tilde{h} \epsilon, (1-(\partial_{\tilde{r}} \tilde{h})^2 \epsilon^2/2)]= -\frac{\epsilon}{\tilde{r}}\partial_{\tilde{r}} (\tilde{r} \partial_{\tilde{r}} \tilde{h} ).
\end{align}
Assuming that the nematic tensor is homogeneous across the thickness and considering an aster-like structure for the microtubules, the first-order force balances in the radial direction and perpendicular to the substrate are given, respectively, by:
\begin{align}
	&
	\hat{r}: -\epsilon^{-2}[\partial_{\tilde{r}}( \tilde{p}+\tilde{\zeta} Q_{rr})-\mu \partial^2_{\tilde{z}} \tilde{v}_r+\frac{\tilde{\zeta}}{\tilde{r}}(Q_{rr}-Q_{\theta \theta})]=0,\label{eq178}\\
	& \hat{z}: -\epsilon^{-3}\partial_{\tilde{z}} (\tilde{p}+\tilde{\zeta} Q_{zz})=0,\label{eq188}
\end{align}
and continuity equation reads
\begin{align}
	&   \frac{1}{\tilde{r}}\partial_{\tilde{r}} (\tilde{r} \tilde{v}_r)+\partial_{\tilde{z}} \tilde{v}_z=0.
\end{align}
Finally, to first order in \( \epsilon \), the boundary conditions can be written as:
\begin{align}
	&  \tilde{v}_z(z=0)=\tilde{v}_r(z=0)=0, \label{noslip1}\\&
	[\tilde{v}_z -\tilde{v}_r \partial_{\tilde{r}} \tilde{h} -\partial_{\tilde{t}} \tilde{h}]_{(z=h)}=\tilde{J},\label{kinematicb}\\&
	\textbf{t}\cdot \boldsymbol{\sigma} \cdot\textbf{n}\sim \epsilon^{-1} \mu [ \partial_{\tilde{z}} \tilde{v}_r ]_{(z=h)}+\epsilon^{-1}\partial_{\tilde{r}} \tilde{h} \tilde{\zeta} (Q_{rr}-Q_{zz})=\partial_{\tilde{r}} \gamma +(\partial_{\tilde{r}} \tilde{h}) \epsilon \partial_{\tilde{z}} \gamma ,\label{stressgama}\\&
	\textbf{n}\cdot \boldsymbol{\sigma} \cdot\textbf{n}\sim  [-\epsilon^{-2}(\tilde{p}+\tilde{\zeta} Q_{zz})=\tilde{p}_a \epsilon^{-2}+\frac{\epsilon}{\tilde{r}} \partial_{\tilde{r}} (\tilde{r} \partial_{\tilde{r}} \tilde{h})\gamma]_{(z=h)},\label{stressgamb}
\end{align}
where $\tilde{p}_a$ and $\tilde{\gamma}$ show the dimensionless air pressure outside the droplet, and  surface tension coefficient, respectively.
Equations~\ref{noslip1} and~\ref{kinematicb} correspond to the no-slip boundary condition at the substrate and the kinematic boundary condition at the interface, respectively. Equations~\ref{stressgama} and~\ref{stressgamb} correspond to the stress boundary conditions at the interface.

Assuming that the activity and the nematic tensor are homogeneous across the thickness, the force balance perpendicular to the substrate (Eq. \ref{eq188}) yields \( \tilde{p} = \tilde{P}(r) - \tilde{\zeta} Q_{zz}(r) \).  Integrating the force balance equation in the radial direction (Eq. \ref{eq178}) over the thickness we obtain:
\begin{align}
	\mu \partial_{\tilde{z}} \tilde{v}_r (Z)= f(r)+Z \partial_{\tilde{r}} (\tilde{p}+\tilde{\zeta} Q_{rr}) +\frac{Z \tilde{\zeta}}{\tilde{r}}(Q_{rr}-Q_{\theta \theta}).
\end{align}
This can be integrated once more, and using the no-slip boundary condition on the substrate, the radial velocity can then be written as
\begin{align}
	\mu \tilde{v}_r (Z)= Z f(r)+Z^2 \partial_{\tilde{r}} (\tilde{p}+\tilde{\zeta} Q_{rr})/2 +\frac{Z^2 \tilde{\zeta}}{2\tilde{r}}(Q_{rr}-Q_{\theta \theta}).\label{eqmuz}
\end{align}
To determine the unknown function \( f(r) \), we apply the stress boundary conditions in the presence of surface tension. In order for the surface tension \( \gamma \) to influence the boundary condition—and, consequently, the radial velocity—we consider the scaling \( \gamma \sim \epsilon^{-1} \tilde{\gamma} \). The stress boundary condition then becomes:
\begin{align}
	&  \mu [ \partial_{\tilde{z}} \tilde{v}_r ]_{(z=h)}+\partial_{\tilde{r}} \tilde{h} \tilde{\zeta} (Q_{rr}-Q_{zz})=\partial_{\tilde{r}} {\tilde{\gamma}} ,\label{bcst}\\&
	-(\tilde{p}+\tilde{\zeta} Q_{zz})=\tilde{p}_a.\label{presex}
\end{align}
The functional form of \( f(r) \) can then be found using Eqs.~\ref{bcst} and \ref{eqmuz} as 
\begin{align}
	f(r)=\partial_{\tilde{r}} \tilde{h} \tilde{\zeta} (-Q_{rr}+Q_{zz})+ \partial_{\tilde{r}} {\tilde{\gamma}}  
	-\tilde{h} \partial_{\tilde{r}} (\tilde{p}+\tilde{\zeta} Q_{rr}) +\frac{\tilde{h} \tilde{\zeta}}{\tilde{r}}(Q_{\theta \theta}-Q_{rr}). 
\end{align}
The radial velocity can then be written as:
\begin{align}
	&\mu \tilde{v}_r (Z)= Z (\partial_{\tilde{r}} \tilde{h} \tilde{\zeta}(-Q_{rr}+Q_{zz})+ \partial_{\tilde{r}} {\tilde{\gamma}}  
	-\tilde{h} \partial_{\tilde{r}} (\tilde{p}+\tilde{\zeta} Q_{rr}) +\frac{\tilde{h} \tilde{\zeta}}{\tilde{r}}(Q_{\theta \theta}-Q_{rr}))\\ \nonumber &+\frac{Z^2}{2} \partial_{\tilde{r}} (\tilde{p}+\tilde{\zeta} Q_{rr}) -\frac{Z^2 \tilde{\zeta}}{2\tilde{r}}(Q_{\theta \theta}-Q_{rr}).
\end{align}
Averaging the radial velocity across the thickness, we then find
\begin{align}
	\langle \tilde{v}_r \rangle = \frac{1}{ \tilde{h}}\int_0^{\tilde{h}} \tilde{v}_r d\tilde{z} = \frac{\tilde{h}}{6\tilde{r}\mu} (3 \tilde{r} f(r)+\tilde{h} \tilde{r} \partial_{\tilde{r}} \tilde{p} + \tilde{\zeta} \tilde{h}(Q_{rr}-Q_{tt}+ \tilde{r} \partial_{\tilde{r}} Q_{rr})).\label{avrad}
\end{align}
The component of the velocity perpendicular to the substrate can then be obtained by integrating the continuity equation:
\begin{align}
	\tilde{v}_z= -\int \partial_{\tilde{r}} (\tilde{r} \tilde{v}_r)/\tilde{r} d\tilde{z} .
\end{align}
From the boundary condition in Eq.~\ref{presex} and the fact that the pressure is constant across the thickness, we find
\begin{align}
	\tilde{p}=-\tilde{p}_{air}-\tilde{\zeta} Q_{zz}.\label{pre}
\end{align}
Substituting the pressure into the average radial velocity (Eq.~\ref{avrad}) for the radial orientation of microtubules ($Q_{\theta z} = Q_{r z} = Q_{r \theta} = 0$, $Q_{z z} = Q_{\theta \theta} = -S/3$, and $Q_{r r} = 2S/3$), and assuming that the magnitude of the nematic order is constant in the region containing the radial orientation, the average radial velocity can then be written as
\begin{align}
	\langle \tilde{v}_r \rangle =-\frac{\tilde{\zeta} \tilde{h} S \left(3 \tilde{r} \partial_{\tilde{r}}\tilde{h}+2 \tilde{h}\right)}{6 \tilde{r} \mu}+
	\frac{\tilde{h} \partial_{\tilde{r}} \tilde{\gamma}}{2 \mu}. \label{finalav}
\end{align}
The dynamics of the droplet thickness can be found using the kinematic bc, radial velocity 
$\tilde{v}_r$ and velocity perpendicular to the substrate $\tilde{v}_z$ at $\tilde{z}=\tilde{h}$, and pressure (Eq. \ref{pre}). This leads to 
\begin{align}
	\partial_{\tilde{t}} \tilde{h} = -\tilde{J}+\frac{1}{\mu \tilde{r}}\partial_{\tilde{r}} [-\tilde{r}\frac{\tilde{h}^3}{3} \partial_{\tilde{r}} \tilde{p}_a+\frac{\tilde{r} \tilde{\zeta} \tilde{h}^3}{3} \partial_{\tilde{r}} (Q_{rr}-Q_{zz})-\tilde{r} \tilde{h}^2 \frac{\partial_{\tilde{r}} \gamma}{2 }+\frac{\tilde{\zeta} \tilde{h}^2}{6}( 3 \tilde{r}(Q_{rr}-Q_{zz}) \partial_{\tilde{r}} \tilde{h} +2 \tilde{h} (Q_{rr}-Q_{\theta \theta}))].
\end{align}
Substituting the orientation of the microtubules in the nematic order parameter $\textbf{Q}$ we find
\begin{align}
	\partial_{\tilde{t}} \tilde{h} = -\tilde{J}+\frac{1}{\mu \tilde{r}}\partial_{\tilde{r}} [-\tilde{r}\frac{\tilde{h}^3}{3} \partial_{\tilde{r}} \tilde{p}_a+\frac{\tilde{r} \tilde{\zeta} \tilde{h}^3}{3} \partial_{\tilde{r}} S-\tilde{r} \tilde{h}^2 \frac{\partial_{\tilde{r}} \tilde{\gamma}}{2 }+\frac{\tilde{\zeta} \tilde{h}^2}{6}( 3 \tilde{r} S \partial_{\tilde{r}} \tilde{h} +2 \tilde{h} S)].
\end{align}
Assuming that the variation of the order parameter $S$ and the thickness is negligible in the area we are interested in, this simplifies to
\begin{align}
	\partial_{\tilde{t}} \tilde{h} = -\tilde{J}+\frac{1}{\mu \tilde{r}}\partial_{\tilde{r}} [-\tilde{r} \tilde{h}^2 \frac{\partial_{\tilde{r}} \tilde{\gamma}}{2 }+\frac{\tilde{\zeta} \tilde{h}^3 S}{3}].\label{eqdth}
\end{align}
Eq. \ref{eqdth} introduces a Marangoni timescale $\tau_M\sim \mu \tilde{R}^2/(\tilde{h} |\Delta \tilde{\gamma}|)$, where $|\Delta \tilde{\gamma}|$ is the change in the surface tension from the contactline to the center of the droplet, an active timescale $\tau_a\sim \mu \tilde{R}^2/(\tilde{\zeta} \tilde{h}^2)$, and an evaporaton timescale $\tau_e\sim\tilde{h}/\tilde{J}$. 
\newpage
\section*{Supplementary Figures}

\subsection*{The Role of Functionalized Substrate}
\label{functionalisedsub}
\begin{figure}[h!]
	\begin{center}
		\includegraphics[width=0.8\textwidth]{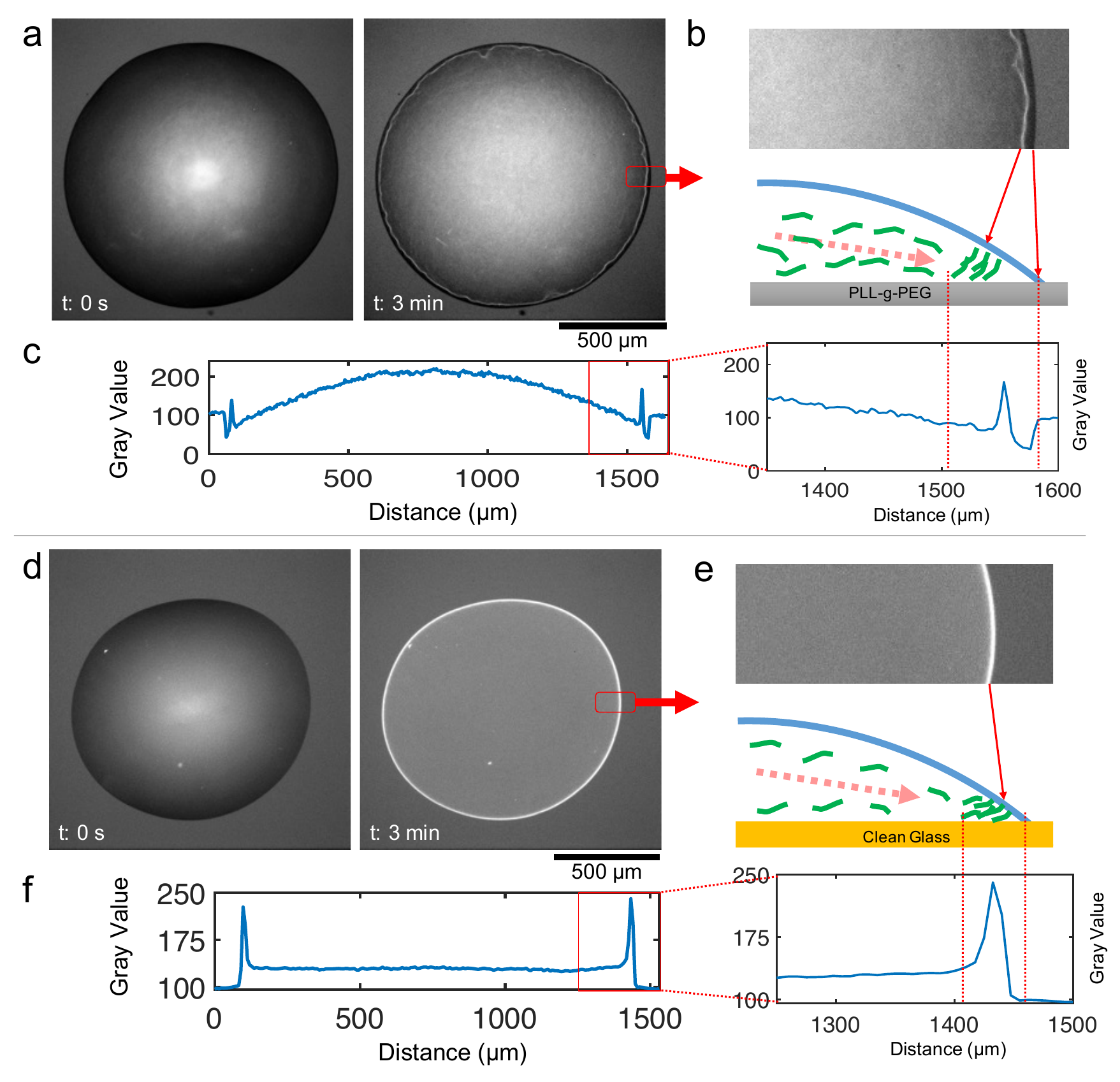}
	\end{center}
	\caption{Comparison of the classical coffee stain formation by passive microtubules in droplets without PEG, on functionalized and non-functionalized glass slides. a) -- c) Micrographs of the evaporating droplet on PLL-g-PEG functionalized glass surface and schematics showing the coffee stain of microtubules on functionalized glass slide. Gray value measurements indicate that the functionalised substrate prevents the microtubules from accumulating at the contact line; instead, their concentration is higher at some distance from it.  In the presence of a functionalised substrate, the distribution of the microtubules inside the droplet is non-uniform and peaks at the center. d) -- f) Micrographs of the evaporating droplet on non-functionalized glass surface and  schematics showing a profile view of the coffee stain of micortubule on non-functionalized glass slide. Gray value measurements indicate the accumulation of the filaments at the exact contact line  and the homogeneous distribution of the microtubules inside the droplet.
	}
	\label{fig:fig5}
\end{figure}
For all the experiments presented in this study, we used glass substrates coated with PLL-g-PEG. This functionalization reduced protein adsorption on the substrate, such as the adhesion of microtubule filaments and motor proteins. 
To understand how reducing protein–substrate interactions might affect the dynamics of the system and its behavior at the droplet’s contact line, we analyzed the evaporation of a passive droplet without PEG on both functionalized and non-functionalized substrates (clean glass; see Materials and Methods section in the main manuscript for more details). PEG was excluded to prevent Marangoni flow, allowing us to study the interaction of microtubules under purely capillary-driven flow.
We compared the initial and final stages of evaporating droplets on two types of surfaces: functionalized (Figure~\ref{fig:fig5}(a)) and non-functionalized (Figure~\ref{fig:fig5}(d)) substrates. At the onset of the experiments, the two droplets exhibited a similar appearance (Figure~\ref{fig:fig5}(a) and (d), $t = 0$~s). After 3 minutes of evaporation, the effects of capillary flow on the passive microtubule network were revealed by measuring the fluorescence intensity of the deposition at the droplet contact line.\\
On functionalized glass, we observed a filament deposition that did not overlap with the droplet contact line (Figure~\ref{fig:fig5}(a), $t = 3$~min). Specifically, a gap in fluorescence intensity was visible between the network and the contact line (Figure~\ref{fig:fig5}(b) and (c)) likely caused by the enhanced spreading of the droplet and the repulsion of biopolymers from the functionalized surface.
\begin{figure}[ht!]
	\begin{center}
		\includegraphics[width=\textwidth]{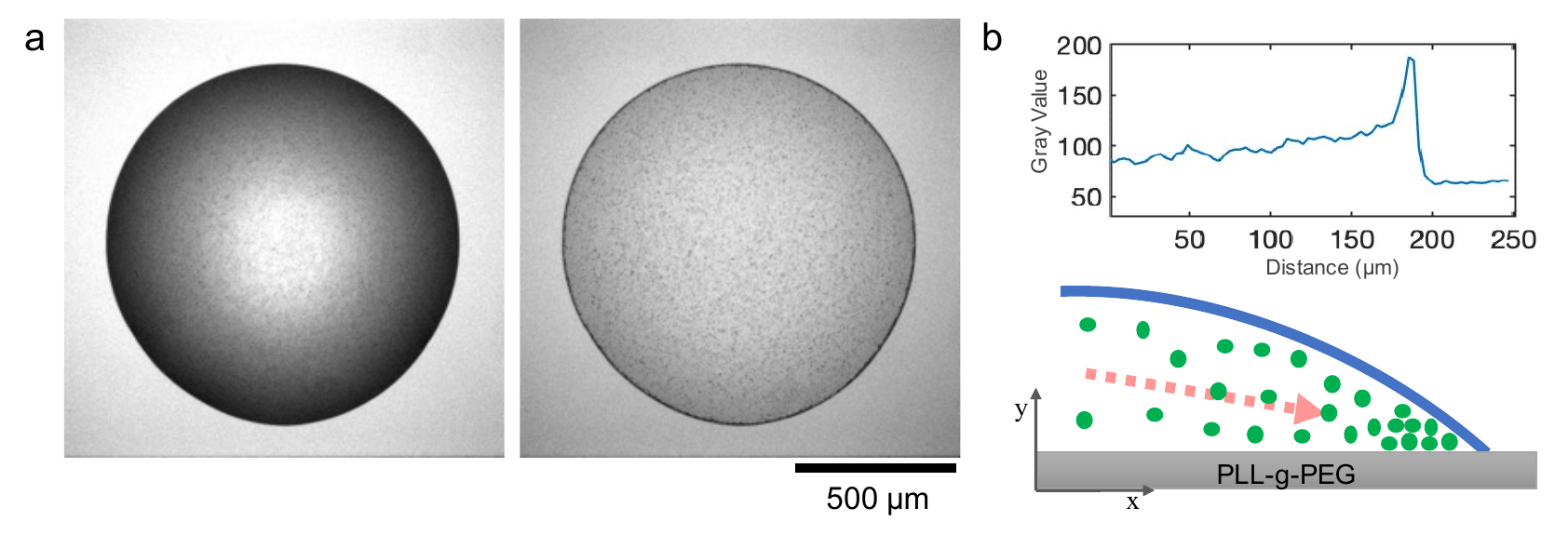}
	\end{center}
	\caption{Formation of coffee ring by spherical particles —much larger than microtubules and insensitive to the repulsive effects of the functionalization— on functionalized glass substrate. a) Wet and dried stages of an evaporating droplet containing buffer solution and spherical particles (980 nm in diamater). The coffee ring can be easily seen at the contact line of the droplet. b) Grey value intensity measurement showing the accumulation of particles at the contact line. }
	\label{sup.fig:S5}
\end{figure}
To test this hypothesis that the functionalisation repeals the microtubules, we repeated the experiments using spherical particles (950~nm in diameter) instead of proteins (microtubules). An evaporating droplet containing spherical particles—much larger than microtubules and insensitive to the repulsive effects of the functionalization—accumulated precisely at the contact line in both cases (see Figure~\ref{sup.fig:S5}).\\
When the biopolymers were deposited on a non-functionalized substrate, a clear ring of microtubules was observed immediately at the droplet’s contact line (Figure~\ref{fig:fig5}(d)). Fluorescence intensity analysis showed no gap between the microtubules and the droplet contact line (Figure~\ref{fig:fig5}(e)), reminiscent of the coffee-ring effect. Furthermore, a uniform fluorescence intensity across the entire droplet footprint emerged after evaporation, corresponding to homogeneous deposition of microtubules on the surface (Figure~\ref{fig:fig5}(f)).
\newpage
\subsection*{The effect of the molecular weight of PEG}

We measured the aggregation ratio $A(t)/A_0$ in zone II for networks formed using PEG with three different molecular weights, where $A(t)$ is the aggregated network area tracked at time $t$, and $A_0$ is the area at $t = 0$~s. Figure S3(d) shows a positive correlation between the aggregation ratio and PEG molecular weight: the 6~kDa PEG network achieved an aggregation of approximately 10\%, while the 20~kDa and 100~kDa PEG networks exhibited stronger aggregation around 20\%. We attributed this effect to a stronger Marangoni flow within the droplet at higher PEG molecular weights. 
\begin{figure}[h!]
	\begin{center}
		\includegraphics[width=\textwidth]{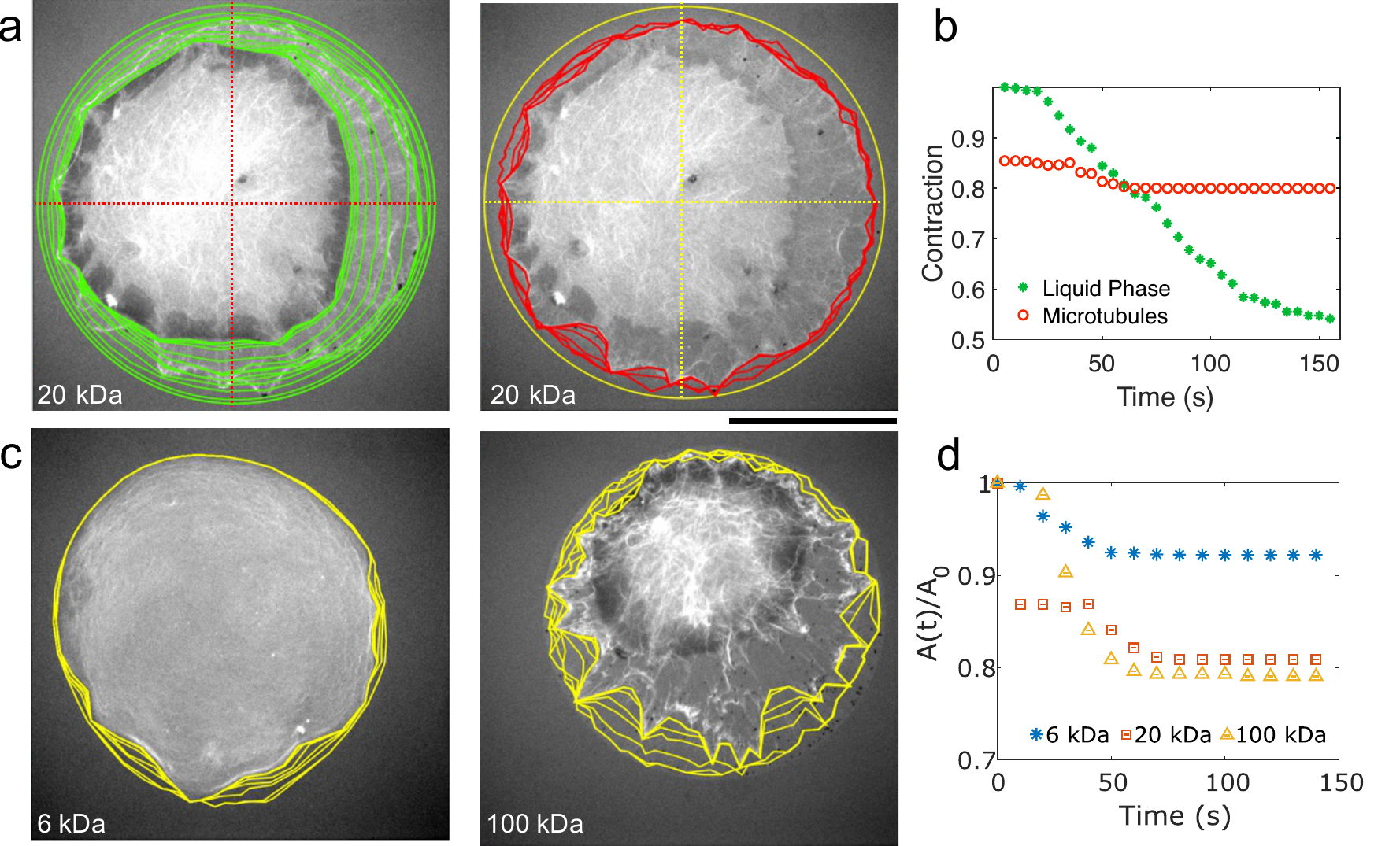}
	\end{center}
	\caption{Aggregation behaviour of passive microtubule network inside an evaporating droplet in the absence of motor proteins and including PEG at different molecular weights. a) Micrographs of the evaporating droplet containing PEG with 20 kDa molecular weight over the course of 5 minutes. b) Quantitative analysis of the contraction over time of the liquid phase containing the internal dense region (in green) and of the network left behind (in red). c) Contraction of droplets with passive microtubule network and PEG with 6 kDa (left) and 100 kDa (right) molecular weights. d) Area of the liquid phase of the evaporating droplet over time for mixtures with PEG with different molecular weights (6, 20, and 100 kDa). The contraction is defined as $A(t)/A_{0}$, with $A(t)$ being the area of the liquid phase over time and $A_{0}$ the initial area of the droplet.}
	\label{fig:fig6}
\end{figure}

\subsection*{Measurements of contraction and expansion of the active droplet}

Active evaporating droplet underwent shrinking and spreading by displaying a distinctive pattern described in the main text. The coupling between the intrinsic active stress of the biological network and the shear stress by Marangoni flow generated two distinct zones: an annulus with radial arrangement of microtubule bundles (Zone I) surrounding a zone in which the biopolymers form an isotropic network (Zone II). By tracking the movement of the contact line and zone II, we observed that they moved inward at a similar velocity during the shrinking stage. 
However, the velocity showed a different trend over the spreading stage, with zone II moving at a greater speed compared to the contact line. 

\begin{figure}[ht!]
	\begin{center}
		\includegraphics[width=0.9\textwidth]{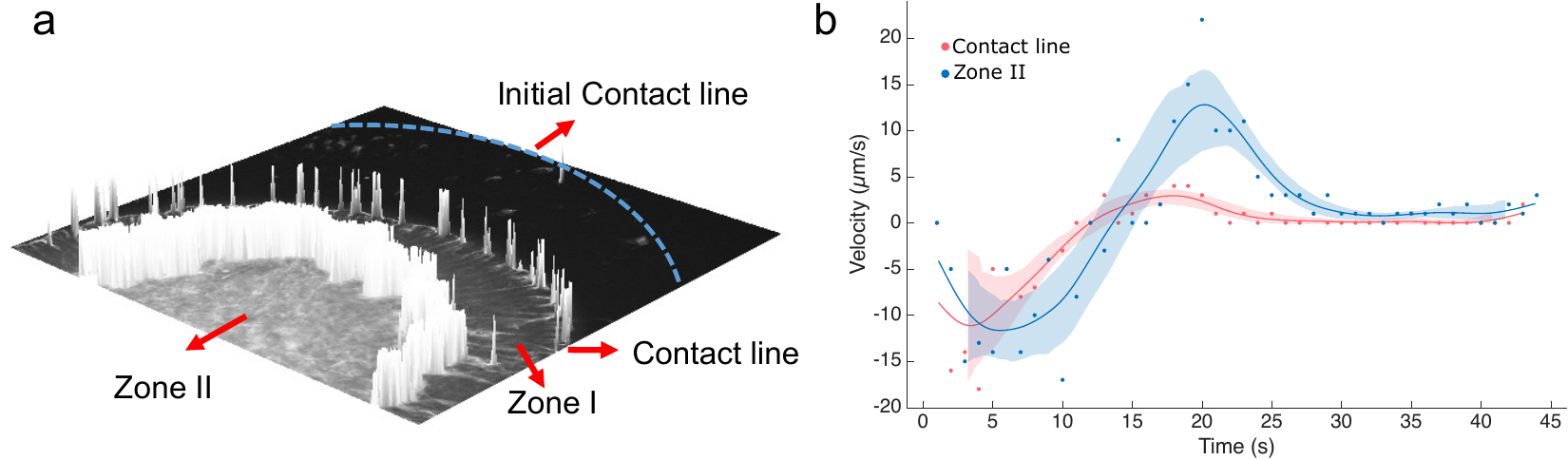}
	\end{center}
	\caption{a) Representation of the 3D ligth profile of the pattern formed in the contracting active network during shrinking (by 3D Surface plot plugin of ImageJ). b) Red and blue show  velocity of the contact line and  network edge (Zone II), respectively.}
	\label{fig:S4}
\end{figure}
\newpage
\paragraph{Drop shape analysis}
By using a Drop Shape Analyser (DSA100M, Kr\"uss, Germany) we measure the volume of droplets with a volume varying between 0.4 $\mu$L and 0.7  $\mu$L deposited on a functionalized glass cover slip.
We analysed samples containing the following constituents: (i) only buffer, (ii) buffer with PEG, (iii) buffer and microtubules, (iv) buffer, PEG, motor proteins, and without ATP.
\begin{figure}[ht!]
	\begin{center}
		\includegraphics[width=0.6\textwidth]{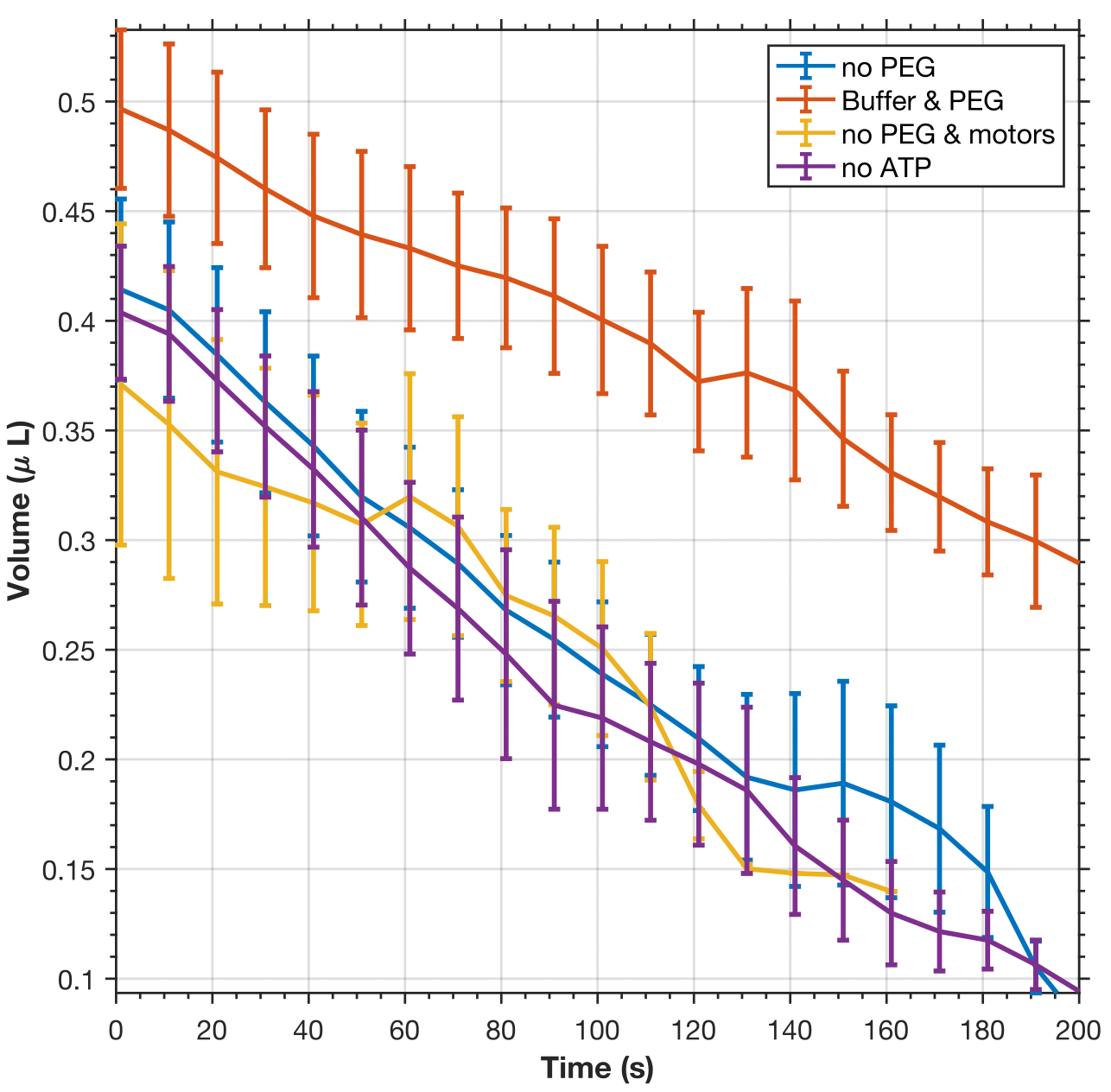}
	\end{center}
	\caption{Volume change of the droplet under different experimental perturbations. Data points and standard deviation were averaged over $20$ s intervals to smooth the data. For the cases (no PEG, Buffer \& PEG, no PEG \& motors, no ATP), the averages and standard deviations are computed from (5, 4, 5, 8) experiments, respectively.
		\label{fig:S5}}
\end{figure}
\begin{figure}[ht!]
	\begin{center}
		\includegraphics[width=0.7\textwidth]{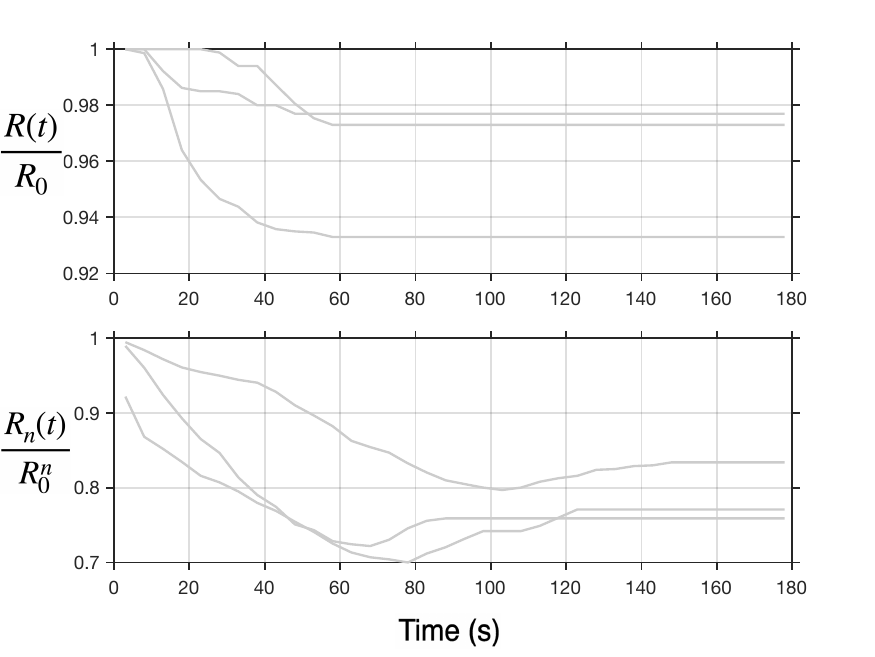}
	\end{center}
	\caption{Dynamics of the evaporating droplet without PEG. The top (bottom) panel shows the dynamics of the contact line (network) in three different experiments. The network first shrinks and then slightly expands. This expansion is evident in the individual experiments. Here, $R_0$ ($R_0^{n}$) denotes the initial radius of the droplet (network), and $R(t)$ ($R_n(t)$) denotes the radius of the droplet (network) over time.
	}
	
	\label{fig:S6}
\end{figure}

\newpage

% *** END NOTES ***
%\subsection*{End Notes}
\subsection*{Acknowledgements}

V.N., O.R-S., S.K. and I.G. acknowledge support from the Max Planck Society. 
V.N. and I.G. acknowledge support form the European Union's Horizon 2020 research and innovation programme under the Marie Sk\l odowska-Curie grant
agreement MAMI No. 766007. I.G. acknowledges support from the Volkswagen Stiftung ("Experiment!").

\subsection*{Author Contributions}
V.N. and I.G. designed all experiments. They performed the experiments presented in the manuscript, except for the micro-PIV and surface-tension measurements, which were carried out by S.K. and O.R.-S., and the evaporation experiments at increased salt concentration, which were performed by S.A. V.N., M.R.N., and I.G. analysed the data. M.R.N. and L.M. developed the theoretical model. V.N., I.G., M.R.N., and L.M. wrote the manuscript. All authors reviewed and edited the manuscript.

\subsection*{Declaration of Interests}
The authors declare no competing interests.
\end{document}